# The homology of defective crystal lattices and the continuum limit

D. H. Delphenich[†]
403 N. Third St., Lindsborg, KS 67456 USA

The problem of extending fields that are defined on lattices to fields defined on the continua that they become in the continuum limit is basically one of continuous extension from the 0-skeleton of a simplicial complex to its higher-dimensional skeletons.  If the lattice in question has defects, as well as the order parameter space of the field, then this process might be obstructed by characteristic cohomology classes on the lattice with values in the homotopy groups of the order parameter space.  The examples from solid-state physics that are discussed are quantum spin fields on planar lattices with point defects or orientable space lattices, vorticial flows or director fields on lattices with dislocations or disclinations, and monopole fields on lattices with point defects.

**Contents**

## 1    Introduction

The crystal lattice plays roughly the same role in solid-state physics that linear spaces do in mathematics (some excellent references on the physics of crystal lattices are [**1-4**]). In either case, one is dealing with an idealization of what one finds in the realm of natural phenomena, although it is still a useful generalization in that the phenomena that one can describe by means of it are still approximately valid within some regime of system parameters. In both cases, one then proceeds into the more realistic natural phenomena by corrupting the idealization with successive levels of defects of manageable types. For instance, Hooke's law can be corrupted by a cubic term in displacement to produce the anharmonic oscillator or the linear law can be replaced with a sinusoidal function to produce the physical pendulum.

---

[†] E-mail: david_delphenich@yahoo.com

The ideal crystal lattice represents a high degree of symmetry as a geometrical structure and, as a result, one finds that the dynamics and physical properties of crystal lattices and their excitations can involve all of the main branches of pure mathematics. In particular, algebra – and mostly the theory of finite groups − bears upon the very definition of the symmetries of a crystal lattice, which then influence the physical properties of the material in a fundamental way. Because the groups in question act as transformation groups by way of symmetries, geometry plays a fundamental role, especially when one introduces dislocations and disclinations into the lattice. Since a lattice is associated with an elaborate system of polyhedra, one can also naturally consider the topological aspects of the lattice by using the methods of simplicial homology, which we will be the focus of the present study. Moreover, the order parameters that one introduces on the lattice often take their values in order parameter spaces that have non-trivial topological properties in their own right. Finally, many of the questions that one asks about crystalline matter are concerned with their physical properties and excitations. Such problems introduce fields of various types, such as tensor fields, wavefunctions, and order parameters that might obey systems of differential equations, and one is ultimately led to the introduction of mathematical analysis into the study, as well. Fourier analysis is a particularly powerful tool in the experimental determination of the structure of crystal lattices.

The subject of the present study is the role of the topology of lattices that contain defects in obstructing the extension of fields that are defined on lattices to fields that are defined on the spaces that these lattices "occupy" when one passes to the continuum limit. One finds that this problem is directly addressable as an elementary problem in the well-established branch of topology that is concerned with topological obstructions to the continuous extension of continuous functions defined on subsets of topological spaces to functions defined on the entire space. Furthermore, it suggests various practical interpretations in terms of familiar models of solid-state and condensed-matter physics.

By now, the role of topology in condensed matter physics has been primarily confined to the study of topological defects in ordered media [**4-14**]. In particular, one mostly considers the homotopy groups of the order parameter space in various dimensions. By comparison, the topology of the space on which the order parameter is defined is usually assumed to be something topologically elementary, such as a vector space or its one-point compactification into a sphere. As a result, the homotopy classes of order parameters that one deals with can be represented as elements of the homotopy groups of the space in which the order fields take their values.

However, solid-state physics also considers defects in the crystal lattice, which are distinct from the topological defects that originate in the homotopy type of the order parameter space. The fact that lattice defects are also a type of topological defect that leads naturally into the study of homology, rather than homotopy, is rarely discussed. When one defines an order field on a defective lattice that takes its values in an order parameter space with non-trivial homotopy, it is then just as natural to define (co-) homology modules for the lattice with coefficients in the homotopy groups of the order parameter space. Furthermore, that is precisely where one finds the obstruction cocyles that tell one whether it is possible to find a continuous extension of an order field on the lattice to an order field on the space it occupies when one passes to the continuum limit.

Actually, the branch of topology known as obstruction theory usually gets more application in the context of extending partial sections of fiber bundles to global sections [**15**]. Although some order fields fall within this purview, when one is concerned with lattices in $\mathbf{R}^n$, whose tangent bundle is trivial, many of the bundle-theoretic aspects of the problem introduce gratuitous generality. Hence, since the methods of homology and obstructions are apparently not common knowledge to the solid-state and condensed matter community, we shall first introduce them in an elementary setting that includes much of the familiar phenomena, and then treat the introduction of non-triviality into the bundle as a further topic to be addressed by a later study.

In the next section of this paper, we define lattices in $\mathbf{R}^n$ in a manner that leads naturally to the usual definitions of solid-state physics, as well as those of simplicial homology. We then briefly review the most common types of lattice defects that solid-state physics deals with and show how they affect the homology of the lattice. In Section 3, we then review the notions of order fields and topological defects in the order parameter spaces, as they are usually considered. In Section 4, we discuss the passage from a crystal lattice to a continuum and the extension of fields defined on lattices to fields defined on continua. Since this suggests the classic formulation of the most elementary problem of obstruction theory, in Section 5 we then recall the basic algorithm that one uses for solving it and apply it to various special cases of interest to solid-state physics.

## 2   Lattices in $\mathbf{R}^n$

Although it is straightforward to mathematically generalize the basic definition of a lattice in $\mathbf{R}^n$ to a lattice in a more topologically interesting manifold, and is probably unavoidable if one is to address "space-time defects [**12-14**]" nevertheless, in the spirit of Occam's Razor, which is one of the foundations of the scientific method, we shall first introduce only as much mathematical generality as is necessary to pose the problem at hand, which only involves lattices of a more prosaic sort, for which the topology is carried by the lattice defects, and not also the space in which they are embedded. The role of the topology of the ambient space can then be introduced later when one wishes to consider its effect on the rest of the theory.

Since there is another usage of the term "lattice" that is quite established in mathematics (for instance, as in [**16**]), namely, a partially-ordered set $(S, \leq)$ that is given two binary operations in the form of glb (greatest lower bound) and lub (least upper bound), we first point out that the "lattice" $\mathbf{Z}^m$, which consists of all ordered $m$-tuples $I = (i_1, \ldots, i_m)$ of integers, and which can also be regarded as the set of all *multi-indices*, can be given a partial ordering that makes the definition of a lattice quite natural. It is simply the "product ordering" that it inherits from the total ordering one defines on $\mathbf{Z}$; that is, $I \leq J$ iff $i_k \leq j_k$ for all $k = 1, \ldots, m$. One then sees that this ordering is not a total ordering anymore, since some *m*-tuples cannot be compared to each other; for instance, when some pairs of indices are greater and others are lesser. The natural definitions of glb and lub are also inherited from the concepts of minimum and maximum from the total ordering on **Z**:

$$\text{glb}(I, J) = (\min\{i_1, j_1\}, \ldots, \min\{i_n, j_m\})$$
$$\text{lub}(I, J) = (\max\{i_1, j_1\}, \ldots, \max\{i_n, j_m\}).$$

We can illustrate this schematically as follows:

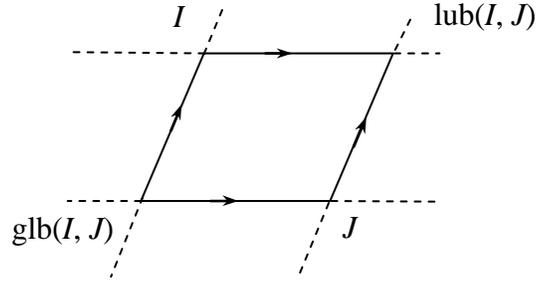

Figure 1. Schematic depiction of a lattice structure.

Our first definition of a lattice in $\mathbf{R}^n$ is more general than we shall need for the remainder of the study, but still quite useful. Let *S* be a finite (but possibly large) subset of $\mathbf{Z}^m$. We define a *lattice in* $\mathbf{R}^n$ to be a one-to-one map $L: S \to \mathbf{R}^n$, $I \mapsto (x_I^1, \cdots, x_I^n)$. Thus, we are making it into a finite subset of $\mathbf{R}^n$ that is parameterized by a finite set of multi-indices in $\mathbf{Z}^m$. Some of the useful general features of this definition are:

 1. We have yet to impose any actual symmetry considerations on the resulting subset in $\mathbf{R}^n$. Hence, in addition to the ions of a crystal lattice, it can just as well describe the instantaneous state of molecules in a fluid or a liquid crystal, the ions of a plasma, the vortices of an Abrikosov flux lattice, or the finite meshes used in the numerical models for systems of differential equations.

 2. It is not necessary to enumerate the elements of the lattice into a total ordering; indeed, such an ordering seems rather arbitrary and unnatural when the lattice is not one-dimensional.

 3. One can consider most of the common examples from solid-state physics by setting $n = 1, 2, 3$, and introduce special relativistic considerations by going to $n = 4$.

 4. We are leaving open the possibility that the dimension of the lattice is strictly less than the dimension of the space that it lives in. (This can be useful when one considers the motion of strings or membranes in higher-dimensional spaces.)

Note that the finite lattice that we have defined will always have a boundary since *S* will. One simply looks at "maximal chains in *S*" of the partial ordering on $\mathbf{Z}^m$ and defines the boundary $\partial S$ to consist of all their endpoints. However, we shall leave the consideration of the boundary out of the immediate discussion, as it rapidly expands the scope of homology to *relative* homology, and will regard that as an extension of the more elementary analysis of the problem at hand.

The traditional terminology for the points of $L(S)$ is *lattice sites* or *lattice points*. In the case of fluids and plasmas, it is reasonable to identify them with positions of the individual molecules or ions, but for crystal lattices, that does not have to be the case.

### 2.1 Basic notions of crystal lattices

In order to define how crystal lattices differ from the more amorphous kind, one must introduce the kind of homogeneity that comes from group actions. Actually, we shall really be using the multiplicative *semi-group* of **Z**, rather than its additive *group*. This is because **Z** is a sub-ring of the ring of real numbers **R**, which act on **R** by multiplication; hence, so do the integers.

Now, let $\{\mathbf{a}_1, \ldots, \mathbf{a}_m\}$ be an *m-frame* in $\mathbf{R}^n$; i.e., a linearly independent set of $m$ vectors in $\mathbf{R}^n$. The product ring $\mathbf{Z}^m$ acts on $\{\mathbf{a}_1, \ldots, \mathbf{a}_m\}$ in the obvious – viz., componentwise – way. The product of the multi-index $I = (i_1, \ldots, i_m)$ and the $m$-frame $\{\mathbf{a}_1, \ldots, \mathbf{a}_m\}$ is the vector:

$$\mathbf{v}_I = \sum_{k=1}^{m} i_k \mathbf{a}_k . \tag{2.1}$$

Hence, every multi-index $I$ in some subset $S \subset \mathbf{Z}^m$ is associated with a distinct vector $\mathbf{v}_I$ in $\mathbf{R}^n$, so we have defined a lattice in $\mathbf{R}^n$ according to our previous definition. We have also defined what is usually referred to as a *Bravais lattice*.

Such a lattice automatically has $m$ spatial periodicities of period $\| \mathbf{a}_k \|$, where the norm used is simply the Euclidian norm on $\mathbf{R}^n$. Similarly, the additive group **Z** acts on each $\mathbf{a}_k$ by scalar multiplication.

More generally, if $A(n)$ is the *affine group* of $\mathbf{R}^n$, which is then the semi-direct product of $GL(n)$ – i.e., all invertible real $n \times n$ matrices – with the translation group $\mathbf{R}^n$, then $A(n)$ acts on $\mathbf{v} \in \mathbf{R}^n$, in the natural way:

$$(A_j^i, b^i) v^j = A_j^i v^j + b^i , \tag{2.2}$$

as well as on any subset of $\mathbf{R}^n$, such as the Bravais lattice generated by $\{\mathbf{a}_1, \ldots, \mathbf{a}_m\}$, which we shall denote by $L(\mathbf{a})$.

A subgroup $G$ of $A(n)$ is said to *preserve* the lattice $L(\mathbf{a})$ if any time $\mathbf{v} \in L(\mathbf{a})$ and $g \in G$ the vector $g\mathbf{v}$ is also an element of $L(\mathbf{a})$. The largest such subgroup, under the partial ordering of inclusion, is called the *space group* of the lattice. More generally, the *crystallographic groups* are the groups that can be the space group for some Bravais lattice. There are two isomorphism classes of crystallographic groups for linear lattices, seventeen for planar lattices, and 230 for space lattices.

The *point group* at $\mathbf{v} \in L(\mathbf{a})$ is the subgroup of the space group of $L(\mathbf{a})$ that fixes $\mathbf{v}$; that also makes a point group the *isotropy subgroup* of the action of the space group on the lattice. It will then be a finite subgroup of the linear subgroup of $A(n)$, namely, $GL(n)$; usually, it consists of powers of some basic rotation, along with reflections. For planar lattices, the only discrete rotation groups that can appear as point groups have order 1, 2, 3, 4, and 6. Other possibilities, such as orders 5 and 7, assert themselves when one is dealing with "quasi-crystals," which are closely related to Penrose tilings, and whose existence in nature was first established experimentally as recently as 1984.

Although we will not actually need to consider the symmetry of a lattice in what follows, we will use one key notion that it implies: that of the unit cell.

One starts with the fact that the $m$-frame $\{\mathbf{a}_1, \ldots, \mathbf{a}_m\}$ in $\mathbf{R}^m$ can be regarded as spanning an $m$-dimensional parallelepiped by means of all vectors of the form (2.1) when the coefficients $i_k$ now range from 0 to 1 without restriction. The volume of this parallelepiped is then:

$$V_c = \det [\mathbf{a}_1| \ldots | \mathbf{a}_m], \tag{2.3}$$

in which the matrix is defined by the components of the vectors $\mathbf{a}_i$, $i = 1, \ldots, m$ with respect to the canonical frame. This parallelepiped then represents the minimum rectangular volume that can be spanned by the basic frame, so one calls the interior of that region of space a *unit* – or *primitive* – cell of the lattice. By definition, each primitive cell contains only one lattice point.

Another way of defining a primitive cell with volume $V_c$ produces the *Wigner-Seitz* cell. First, one connects a chosen lattice point to all of its nearest neighbors by means of line segments. (The *nearest neighbors* of a given lattice point $I = (i_1, \ldots, i_m)$ are characterized by those points $J$ of the lattice whose coordinates differ from those of $I$ by either $-1$, 0, or $+1$.) One then intersects a normal hyperplane through the midpoint of each such line segment. The smallest-volume polyhedron that these hyperplanes bound is then the Wigner-Seitz cell.

The significance of primitive cells for us is that they give us an indication of what sort of geometric building blocks we should use to define a simplicial complex that would represent the homology of the crystal lattice. Although there are five basic types of planar Bravais lattices and fourteen types of Bravais space lattices, one finds, upon perusing the illustrations (cf., e.g., Kittel [**1**]), that they tend to suggest two natural types of building blocks: parallelepipeds, which one can regard as *cubic simplexes*, and *triangular simplexes,* which we will refer to as *cubes* and *simplexes*, respectively.

If the lattice is of the type that suggests the use of parallelepipeds then a *1-cube* will be a line segment between two distinct lattice sites, a *2-cube* will be the parallelogram spanned by four distinct lattice sites when two pairs of them generate distinct parallel lines, and a *3-cube* is the three-dimensional parallelepiped that is generated by eight distinct lattice sites, two quadruples of which generate distinct parallel planes.

In general, the $m$-frame $\{\mathbf{a}_1, \ldots, \mathbf{a}_m\}$ spans an $m$-dimensional *simplex* by letting the coordinates $x^k$, $k = 1, \ldots, m$ of a point $\mathbf{x} = x^k \mathbf{a}_k$ range from 0 to 1 under the restriction that their sum must always be 1:

$$1 = \sum_{k=1}^{m} x^k . \tag{2.4}$$

One sees that volume of the region that is spanned by an $m$-simplex is one-half the volume of corresponding primitive cell.

Hence, a *1-simplex* is a line segment in $\mathbf{R}^n$ that connects two distinct lattice sites, a *2-simplex* is a triangle spanned by three distinct, non-collinear lattice sites (including its interior points), and a *3-simplex* is a tetrahedron spanned by four non-coplanar lattice sites. One can continue the process into higher dimensions, but for the present purposes that will not be necessary.

One notices that as far as topology is concerned a $k$-simplex and a $k$-cube are the same thing; i.e., they are *homeomorphic*. Moreover, they are both homeomorphic to a closed $k$-dimensional ball $B^k$; generically, we shall refer to all three of them as $k$-cells. They will define the basic building blocks for the homological structure of the lattice. The choice of one representation or the other for a $k$-cell then amounts to whichever representation is most natural to the problem.

### 2.2 Homology of lattices in $\mathbf{R}^n$

The set $S$ can define what one calls the *vertex system* for an *abstract simplicial complex* [**17-19**]; we shall also refer to it as the *0-skeleton* of that complex. The corresponding points $\mathbf{v}_I$ of the lattice $L: S \to \mathbf{R}^n$ then define the *geometric realization* of the vertex system and we denote it by $L_0$.

The *1-simplexes* of this abstract simplicial complex are then a specified subset $S_1$ of $S \times S$; that is, a specified collection of pairs $(I, J)$ of multi-indices in $S$. For an ideal crystal, there are two possibilities for this specified subset: If the lattice suggests cubic simplexes as its building blocks then one includes all pairs $(I, J)$ of vertices such that one pair of lattice coordinates differs by $\pm 1$. We denote this in the usual multi-indicial way by:

$$J = I + 1_k = (i_1, \ldots, i_k + 1, \ldots, i_m) \tag{2.5}$$

Otherwise, when the basic building blocks are triangular simplexes, one includes pairs such that $I$ and $J$ are nearest neighbors in $S$, although not necessarily all such pairs.

We illustrate some of these possibilities for planar lattices:

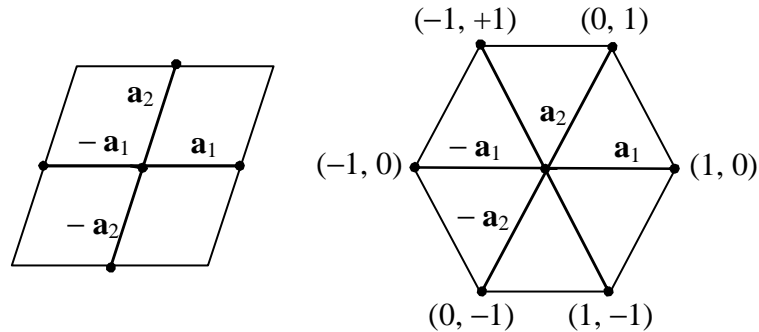

Figure 2. Some possible ways of expressing lattices in terms of elementary cells.

The geometric realization of $S_1$ then consists of all line segments in $\mathbf{R}^n$ that connect the vertex pairs of $L(S)$ when the multi-index pair is in $S_1$; we denote that set by $L_1$. As far as topology is concerned, the use of a straight line segment is not necessary, and one can just as well ise a continuous non-self-intersecting curve.

It might be physically reasonable to say that when lattice sites represent atomic ions the inclusion of the 1-simplex $IJ$ defined by two vertices $I$ and $J$ in $S_1$ would correspond to their being bound by some relevant crystal binding force, such as ionic forces, Van der Waals forces, or covalent bonding.

The *2-simplexes* are a specified subset $S_2$ of either triples $(I, J, K)$ of distinct, non-collinear multi-indices, in the case of hexagonal lattices, or quadruples $(I, J, K, L)$ of distinct multi-indices, two pairs of which generate distinct, parallel lines, in the cubic case. For an ideal lattice, all such triples or quadruples are included in $S_2$. The geometric realization of $S_2$ then consists of all triangles or parallelograms in $\mathbf{R}^n$ that are spanned by the vertices that correspond to the multi-indices of the triples or quadruples of $S_2$, respectively; we denote that set by $L_2$. By the words "triangle" or "parallelogram," we are intending that one includes all of their interior points as surface segments, as well as the points of their edges. One way of defining these interior points is to form all convex combinations $\lambda x^i + (1 - \lambda) y^i$, $0 \leq \lambda \leq 1$, of point pairs $\{x^i, y^i\}$ taken from the edge points.

The *3-simplexes* are then a specified subset $S_3$ of either quadruples of multi-indices that generate tetrahedral in $\mathbf{Z}^m$ or octuples that generated parallelepipeds. Their geometric realizations are the corresponding tetrahedral or parallelepiped regions generated by the convex closure of the corresponding vertices in $\mathbf{R}^n$; we denote that set by $L_3$. Ultimately, one concludes with $L_m$ as the geometric realization of the lattice in $\mathbf{R}^n$, since simplexes of dimension higher than $m$ cannot exist; of course, for us, $m$ will general be 1, 2, or 3.

In general, we shall refer to the simplexes that pertain to the hexagonal lattices – viz., nodes, branches, triangles, and tetrahedral – as *triangular simplexes*, while the one that pertain to cubic lattices – viz., nodes, branches, rectangles, and parallelepipeds – as *cubic simplexes*. In one sense, the triangular simplexes are the most fundamental, since any polyhedral region can ultimately be subdivided into triangular subregions. One of the early advances of simplicial homology was the proof that subdividing a given triangular simplicial complex into a finer complex by means of a "barycentric" subdivision, which adds points at the centers of "mass" of the simplexes in each dimension and connects them with additional simplexes in each dimension, does not affect the homological information that it contains.

Previously, we called $S$ the 0-skeleton of the abstract simplicial complex and $L(S)$ the 0-skeleton of its geometric realization. More generally, the *k-skeleton* $\Sigma_k$ of the abstract simplicial complex is the union $S_0 \cup \ldots \cup S_k$, while that of the geometric realization is $\Lambda_k = L_0 \cup \ldots \cup L_k$. In either case, the entire complex coincides with its $m$-skeleton. One then refers to the dimension of the complex as being the highest dimension of its simplexes.

For the sake of brevity, it is convenient to denote $k$-simplexes as "words" formed from the "letters" of the "alphabet" $S$. For instance, the 1-simplex $(I, J)$ is denoted by $IJ$, the 2-simplex $(I, J, K)$ by $IJK$, and the 3-simplex $(I, J, K, L)$ by $IJKL$, in the event that the unit cell is a tetrahedron; the corresponding expressions for the cubic case are analogous.

The (abstract) *boundary* of a 1-simplex $IJ$ is the set of its vertices: $\partial(IJ) = \{I, J\}$. The boundary of a 2-simplex consists of all of its consecutive vertex pairs – i.e., edges – taken in the sequence defined by the word, and including the last letter followed by the first one.

For the hexagonal lattice:
$$\partial(IJK) = \{IJ, JK, KI\},$$
while for the cubic lattice:

$$\partial(IJKL) = \{IJ, JK, KL, LI\}.$$

The boundary of a 3-simplex is defined differently for the triangular and cubic cases. In the triangular case, one considers its four consecutive vertex triples – or *faces* – when one repeats the first two letters after the last one:

$$\partial(IJKL) = \{IJK, JKL, KLI, LIJ\}.$$

For the cubic case, one must decompose the parallelepiped into its six rectangular faces:

$$\partial(I_1I_2I_3I_4I_5I_6I_7I_8) = \{I_1I_2I_3I_4, I_1I_5I_6I_2, I_1I_5I_8I_4, I_2I_6I_7I_3, I_3I_7I_8I_4, I_5I_6I_7I_8\}.$$

In general, $\partial\sigma_k$ consists of all of its $k-1$-*faces*.

Naturally, one must add the restriction on the definition of an abstract simplicial complex that $L_k$ must always contain the boundaries of all simplexes in $L_{k+1}$.

If one regards all permutations of letters in a word that represents a simplex as equivalent then one thinks of that simplex as being unoriented. Hence, one then thinks of a choice of permutation for the sequence of letters for a $k$-simplex as defining an *orientation* for that $k$-simplex. More generally, one can think of any permutation of the letters of $\sigma_k = IJ…K$ as being one or the other orientation for $\sigma_k$ according to the sign of the permutation (+ = even, − = odd). One arbitrarily assigns a + or − sign to $\sigma_k$ according to which type of permutation of $IJ…K$ defines it. In the case of line segments, one can think of orientation as defining a sense of motion from one vertex to the other, while for a triangle or parallelogram an orientation is a sense of rotation for a circuit around the vertices by way of the edges.

What makes triangular simplexes particularly convenient is that one can obtain the faces of $\sigma_k$ by successively deleting each letter from the word $IJ...K$ and multiplying by the sign of the permutation that takes the resulting word to the oriented word that is included in the set of abstract $k-1$-simplexes. The faces of a cubic simplex are not as concisely described.

A *k-chain* $c_k$ is defined to be a finite formal sum of oriented $k$-simplexes $\sigma_k(i)$, $i = 1, …, N$ with integer coefficients $m_i$:

$$c_k = \sum_{i=1}^{N} m_i \sigma_k(i). \tag{2.6}$$

Although the concept of a formal sum can be made mathematically rigorous, for the sake of computation, it is usually sufficient to treat it as a set of rules for symbol manipulation. Hence, to be consistent, one must have:

$$m\sigma_k = \sigma_k + … + \sigma_k \quad (k \text{ summands}), \qquad (-1)\sigma_k = -\sigma_k. \tag{2.7}$$

As a consequence: $(1)\sigma_k = \sigma_k$.

In order to interpret $(0)\sigma_k$, one must think of multiplication by 0 as eliminating that term from the sum, and the "empty sum" is represented by 0. Since the empty set is a subset of any set, 0 is a *k*-chain for every *k*. Hence, one can define the addition and subtraction of *k*-chains by implicitly including summands of the form $(0)\sigma_k = 0$ if necessary. If $c_k$ is as in (2.6) and $c'_k$ is defined analogously then:

$$c_k + c'_k = \sum_{i=1}^{N}(m_i + m'_i)\sigma_k(i),\tag{2.8}$$

which then makes:

$$c_k - c'_k = c_k + (-1)c'_k, \qquad c_k - c_k = (1-1)c_k = 0.\tag{2.9}$$

The set $C_k(\Sigma_m)$ or $C_k(\Lambda_m)$ of all *k*-chains formed from the *k*-simplexes of either the abstract simplicial complex generated by $S$ or its geometric realization $L$ in $\mathbf{R}^n$, respectively, when given the scalar multiplication by integers (2.7) and the addition (2.8), defines what is called a *free* **Z**-*module* (for more details on modules, see, e.g., MacLane and Birkhoff [**20**]) Such an algebraic structure is essentially a vector space whose vectors are the *k*-chains, but whose scalars come from a ring that is not a field, namely, **Z**; hence, multiplicative inverses to scalars do not usually exist (except for 1).

The *rank* of $C_k(\Sigma_m)$ or $C_k(\Lambda_m)$ is the analogue of the dimension of a vector space and equals the number $r_k$ of *k*-simplexes in $S_k$, in this case. Since the dimensions of $C_k(\Sigma_m)$ or $C_k(\Lambda_m)$ are the same, they are isomorphic as free **Z**-modules, and we shall henceforth deal with only $C_k(\Lambda_m)$. The fact that we are dealing with finite simplicial complexes implies that each $C_k(\Lambda_m)$ has a finite rank, and since it is a free module, one can treat the generating set $\{\sigma_k(i), i = 1, \ldots, r_k\}$ as a generalized basis for the module, although one only forms linear combinations with integer coefficients.

Since non-zero *k*-chains do not exist for $k > m$ one already has $C_k(\Lambda_m) = 0$, $k > m$.

If one now regards an abstract triangular *k*-simplex $\sigma_k = I_0\ldots I_k$ as a *k*-chain then one can associate its abstract boundary, in the sense above, with a *k*–1-chain:

$$\partial_k \sigma_k = \sum_{i=0}^{k}(-1)^i I_0 \cdots \hat{I}_i \cdots I_k,\tag{2.10}$$

in which the caret over $I_i$ means that the letter in question is deleted from the word in that summand. In particular:

$$\partial_1(IJ) = J - I,\tag{2.11}$$
$$\partial_2(IJK) = JK - IK + IJ = IJ + JK + KI,\tag{2.12}$$
$$\partial_3(IJKL) = JKL - IKL + IJL - IJK.\tag{2.13}$$

If one uses cubic *k*-chains instead of triangular ones then the analogue of (2.10) is not as simple to write down, but we can say that if the edges of a square are oriented in a counterclockwise circular manner and the faces of a parallelepiped are oriented positive outward then the analogues of (2.12) and (2.13) are:

$$\partial_2(IJKL) = IJ + JK + KI + KL,\tag{2.14}$$

$$\partial_3(I_1I_2I_3I_4I_5I_6I_7I_8) = I_1I_2I_3I_4 + I_1I_5I_6I_2 + I_1I_5I_8I_4 + I_2I_6I_7I_3 + I_3I_7I_8I_4 + I_5I_6I_7I_8. \quad (2.15)$$

One can then extend this definition of the boundary operator by linearity to a linear operator $\partial_k : C_k(\Lambda_m) \to C_{k-1}(\Lambda_m)$. That is, if $c_k$ is as in (2.6), with $N = r_k$, by allowing some $m_i$ to be zero, then:

$$\partial_k c_k = \sum_{i=1}^{r_k} m_i \, \partial_k \sigma_k(i) = \sum_{i=1}^{r_k} \sum_{j=1}^{r_{k-1}} ([\partial_k]_{ji} m_i) \sigma_{k-1}(j), \quad (2.16)$$

in which $r_{k-1}$ represents the total number of all $k-1$-simplexes.

The integer matrix $[\partial_k]_{ji}$ is $r_{k-1} \times r_k$ and when the simplexes are unoriented, its elements equal 1 if $\sigma_{k-1}(j)$ is a face of $\sigma_k(i)$ and 0 if it is not one of the faces. When the simplexes have been given specific orientations, the 1's become + 1 or – 1, depending upon whether $\pm \sigma_{k-1}(j)$ is an oriented face of $\sigma_k(i)$, respectively. One calls $[\partial_k]_{ji}$ the *incidence matrix* of $\sigma_k(i)$, and it basically represents the matrix of the linear map $\partial_k$ with respect to the bases for the modules $C_k(\Lambda_m)$ and $C_{k-1}(\Lambda_m)$ defined by $\sigma_k(i)$ and $\sigma_{k-1}(j)$. If one lets $\partial_k c_k = \sum_{j=1}^{r_{k-1}} n_j \sigma_{k-1}(j)$ then one can then replace (2.16) with the system of linear equations:

$$n_j = [\partial_k]_{ji} m_i . \quad (2.17)$$

Sometimes the elements of the incidence matrix are expressed as simply the *incidence numbers* $[\sigma_k(i), \sigma_{k-1}(j)]$, especially when one is dealing with modules that are not finitely generated. For instance, a triangulation [1] of a non-compact topological space will not have finitely many simplexes to it, and when one goes to singular homology the number of singular simplexes in each dimension will generally be uncountably infinite. Hence, one of the advantages of specializing the tools of homology to spaces that admit finite triangulations is that one can define bases for the modules of chains and treat the boundary operator as an integer matrix with a finite number of rows and columns.

By definition, the boundary of any 0-simplex, and therefore any 0-chain is always zero; i.e., $\partial_0 : C_0(\Lambda_m) \to 0$.

A physically interesting way of interpreting the boundary of a 1-chain $c_1$ is to think of $c_1$ as representing an electrical network whose branches are the 1-simplexes and whose currents $I_i$ are the coefficients in the formal sum. If one writes $c_1$ in the form $\sum_{i=1}^{r_1} I_i \sigma_1(i)$ ($r_1$ = number of branches in the network) then its boundary is:

$$\partial_1 c_1 = \sum_{i=1}^{r_1} I_i \partial \sigma_1(i) = \sum_{j=1}^{r_0} \left( \sum_{i=1}^{r_1} [\partial_1]_{ji} I_i \right) \sigma_0(j), \quad (2.18)$$

---

[1] We shall use the word "triangulation" to mean a decomposition of a topological space into the geometric realization of an abstract simplicial complex regardless of whether the basic building blocks are triangular or cubic simplexes.

in which $r_0$ represents the number of nodes in the network.

Thus, the coefficient of each node of the network – i.e., each 0-simplex $\sigma_0(j)$ – in $\partial_1 c_1$ is $\sum_{i=1}^{r_1}[\partial_1]_{ji} I_i$, which represents the sum of the (signed) currents in the branches that have that node for a boundary component, and if $\partial c_1$ vanishes then one has a homological statement of Kirchhoff's law of currents: The current in a network is a 1-chain with real coefficients, and in the equilibrium state (constant currents, no accumulation of charge at any node) it has boundary zero [2]; i.e.:

$$\sum_{i=1}^{r_1}[\partial_1]_{ji} I_i = 0 \text{ for each } j. \tag{2.19}$$

A basic property of the boundary operator, which can be verified directly from (2.10), is that its "square" always vanishes:

$$\partial_{k-1} \partial c_k = 0 \qquad (\text{i.e.,} \quad \partial_{k-1}(\partial c_k) = 0 \text{ for any } c_k). \tag{2.20}$$

Since $\partial$ is linear, its kernel (i.e., the set of all $k$-chains with boundary zero) is also a **Z**-module that is a submodule of $C_k(\Lambda_m)$, and we denote this kernel by $Z_k(\Lambda_m)$; an element of this **Z**-module is called a *k-cycle*. Similarly, the image of $\partial$ will also be a **Z**-submodule of $C_{k-1}(\Lambda_m)$ that we denote by $B_{k-1}(\Lambda_m)$; these $k-1$-chains are called *k−1-boundaries*. Note that since there are no $(m+1)$-chains, $B_m(\Lambda_m) = 0$, while all 0-chains are cycles, so $Z_0(\Lambda_m) = C_0(\Lambda_m)$.

From (2.20), one sees that every $k$-boundary is also a $k$-cycle, so $B_k(\Lambda_m)$ is a **Z**-submodule of $Z_k(\Lambda_m)$. The crucial question at the root of homology is whether the converse statement is true. In general, there might be $k$-cycles that do not bound $k+1$-chains. One can think of a triangle or square minus its interior points as examples of 1-cycles (the sequence of oriented edges) that do not bound a 2-chain. Hence, a non-bounding $k$-cycle represents a sort of "$k+1$-dimensional hole" in the space described by a chain complex.

One then defines the quotient module $H_k(\Lambda_m) = Z_k(\Lambda_m) / B_k(\Lambda_m)$ to be the (integer) homology module in dimension $k$ for either the abstract simplicial complex or its geometric realization in $\mathbf{R}^n$. That is, an element of $H_k(\Lambda_m)$ is an equivalence class of $k$-cycles under homology: Two $k$-cycles $z_k$ and $z'_k$ are said to be *homologous* if their difference is the boundary of a $k+1$-chain:

$$z_k \sim z'_k \qquad \text{iff} \qquad z_k - z'_k = \partial c_{k+1} \qquad \text{for some } c_{k+1}. \tag{2.21}$$

One then sees that when a $k$-cycle is a boundary it is "homologous to zero."

In particular, two vertices $v_1$ and $v_2$ are homologous iff there is some connected 1-chain from one to the other, which then amounts to a continuous path in the geometric realization. Hence, the zero-dimensional homology module $H_0(\Lambda_m)$ has a basis that is defined by the path-connected components of $\Lambda_m$, so if $\Lambda_m$ is path-connected then $H_0(\Lambda_m)$ is **Z**.

---

[2] For more discussion of the role of homology in electrical networks, one can confer Lefschetz [21].

Once again, since there are no ($m$+1)-chains but 0, one always has $H_m(\Lambda_m) = Z_m(\Lambda_m)$.

Homology is related to homotopy, but not equivalent. In particular, although two homotopic $k$-chains will be homologous, the converse is not generally true. We illustrate some possibilities for homologies using 1-cycles that we represent as circles in Figure 3.

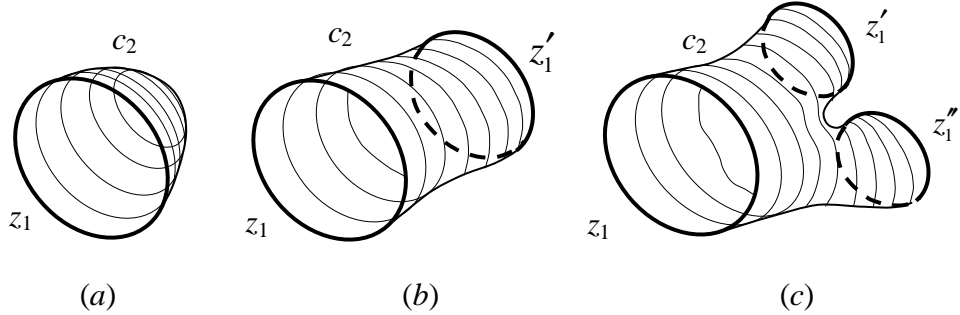

Figure 3. Examples of homologies of 1-cycles.

The ($a$) illustration depicts a 1-cycle $z_1$ that is homologous to 0 ($z_1 = \partial c_2$); when $z_1$ is a loop this also makes it homotopic to a point (i.e., to a constant map). In the ($b$) illustration, we see a case of two homologous 1-cycles that are homotopic in a non-trivial way, and in ($c$) the 1-cycle $z_1$ is homologous to the 1-chain $z_1' + z_1''$, which does not represent a homotopic image of $z_1$.

One finds that the first (i.e., one-dimensional) homology module $H_1(\Lambda_m)$, when regarded as an Abelian group (i.e., ignoring the scalar multiplication), is isomorphic to the "Abelianization" of the fundamental group $\pi_1(\Lambda_m)$; hence, when $\pi_1(\Lambda_m)$ is Abelian to begin with, they are isomorphic. Thus, the first homology module of a figure eight is the same as the homology of two disjoint circles, although their fundamental groups differ. (For more details on the computation of homotopy groups, one might confer Massey [**22**].)

It is important to note that whereas the module $C_k(\Lambda_m)$ was free, so it took the form of $\mathbf{Z}^{r_k}$, where $r_k$ is the number of $k$-simplexes, the same is not true for $H_k(\Lambda_m)$, in general. It will generally be the direct sum of a free module $\mathbf{Z}^{b_k}$, where $b_k$ is called the $k^{th}$ *Betti number* of $\Lambda_m$, and a finite number of finite cyclic groups, which one refers to as the *torsion part* of $H_k(\Lambda_m)$. That means that for some $k$-cycles $z_k$ there will be a non-zero integer $N$ such that $Nz_k = \partial c_{k+1}$ for some $c_{k+1}$. An elementary example of a topological space in which torsion appears in the integer homology is any real projective space $\mathbf{R}P^n$. In fact, when $n > 1$, if one represents $\mathbf{R}P^n$ as the quotient of the $n$-sphere $S^n$ by the identification of antipodal points then the homology modules do not change, except in dimension one, where a torsion summand of $\mathbf{Z}_2$ appears in place of 0. Thus, not only does $\mathbf{R}P^n$ become multiply-connected as a result of the identification, when $n$ is even it also becomes non-orientable, as a manifold; we shall not go into that here, but refer the curious to the chapters in Vick [**23**] or Greenberg [**24**] that are concerned with the topology of manifolds. Note that $\mathbf{R}P^3$ is not only orientable, it is parallelizable, in the sense that it admits a global, continuous field of tangent linear 3-frames.

Generally, the Betti number in each dimension is equal to the number of "holes" of that dimension (plus one). For instance, one can say that an *n*-sphere has an (*n*+1)-dimensional hole in it, unless the interior points are included, in which case, it becomes contractible to a point or homologous to 0.

When one forms the alternating sum of the Betti numbers the resulting integer is called the *Euler-Poincaré characteristic* of the topological space in question:

$$\chi[\Lambda_m] = \sum_{k=0}^{m} (-1)^k b_k .  \qquad (2.22)$$

Interestingly, this number also equals the alternating sum $\sum_{k=0}^{m} (-1)^k r_k$ of the ranks of the modules $C_k(\Lambda_m)$, even though the boundary operator has not been introduced at that point. For instance, in the case of a triangulated compact surface $\Sigma$:

$$\chi[\Sigma] = \# \text{ vertices} - \# \text{ branches} + \# \text{ faces},  \qquad (2.23)$$

which is the form that Euler himself used in his treatment of the Königsberg Bridges problem.

Another important point to comprehend about homology is that the only structure that the simplexes of each dimension contribute to the modules of chains in those dimensions is due to their raw number; i.e., the free $\mathbf{Z}$-module generated by a set of *N* apples is isomorphic to the free $\mathbf{Z}$-module generated by a set of *N* oranges, or anything else. The only way that topology enters into the picture is by way of the definition of the boundary operator itself, which is then tailored to the demands of the specific problem. For instance, we illustrate this process in the elementary cases of simplicial complexes that represent a circle and a disc in Figure 4.

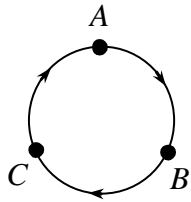   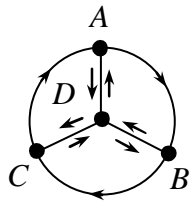

$\partial_1(AB) = B - A$
$\partial_1(BC) = C - B$
$\partial_1(CA) = A - C$
$\partial_1(AB + BC + CA) = 0$

$\partial_2(ABD) = BD - AD + AB$
$\partial_2(BCD) = CD - BD + BC$
$\partial_2(ADC) = DC - AC + AD$
$\partial_2(ABD + BCD + ADC) = AB + BC + CA$

Figure 4. Some elementary triangulations.

If we let $\sigma_0(i)$, $i = 1, 2, 3$ be $A$, $B$, and $C$, resp., while $\sigma_1(j)$, $j = 1, 2, 3$ represent $AB$, $AC$, and $BC$, resp. then the boundary operators $\partial_1$ and $\partial_0$ for the circle have the incidence matrices:

$$[\partial_1]_{ij} = \begin{bmatrix} -1 & 1 & 0 \\ -1 & 0 & 1 \\ 0 & -1 & 1 \end{bmatrix}, \qquad [\partial_0]_i = [0, 0, 0] \tag{2.24}$$

If we let $\sigma_0(i)$, $i = 1, 2, 3, 4$ be $A$, $B$, $C$, and $D$, resp., while $\sigma_1(j)$, $j = 1, \ldots, 6$ represent $AB$, $AC$, $AD$, $BC$, $BD$, $CD$, resp., and $\sigma_2(k)$, $k = 1, 2, 3$, represent $ABC$, $BCD$, $ADC$, resp. then the boundary operators $\partial_1$ and $\partial_0$ for the circle have the incidence matrices:

$$[\partial_2]_{jk} = \begin{bmatrix} 1 & 0 & 0 \\ 0 & 0 & -1 \\ -1 & 0 & 1 \\ 0 & 1 & 0 \\ 1 & -1 & 0 \\ 0 & 1 & -1 \end{bmatrix}, \qquad [\partial_1]_{ij} = \begin{bmatrix} -1 & -1 & -1 & 0 & 0 & 0 \\ 1 & 0 & 0 & -1 & -1 & 0 \\ 0 & 1 & 0 & 1 & 0 & -1 \\ 0 & 0 & 1 & 0 & 1 & 1 \end{bmatrix},$$

$$[\partial_0]_i = [0, 0, 0, 0] \tag{2.25}$$

One can represent all of the compact surfaces without boundary by means of identifications of the edges of a square, which then leads to triangulations of them that allow one to also express their homology modules (see Massey [**22**]).

### 2.3 Lattice defects

Having already introduced more mathematical abstraction than most solid-state physicists feel comfortable with, we now return to the realm of real-world crystal lattices to apply those abstractions to the homology of defective lattices.

If one were dealing with an ideal crystal lattice then every pair of lattice sites would be connected by some 1-chain, every 1-cycle would bound some 2-chain, and every 2-cycle would bound some 3-chain. Hence, the homology modules of an ideal lattice all vanish, making homology a pointless complication to introduce in that context.

However, in the real world crystal lattices are never ideal, but contain lattice defects of various types. The three basic types of lattice defects are point defects, linear defects, and surface defects.

Point defects take the form of lattice vacancies, interstitial inclusions of ions that ordinarily belong to the lattice, and interstitial inclusions or lattice substitutions of impurities. We illustrate these possibilities as follows:

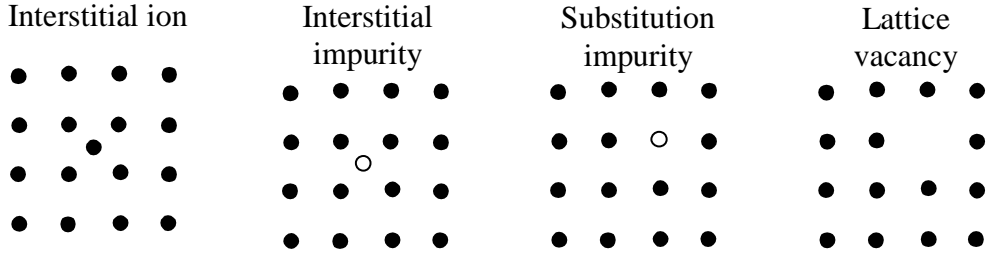

Figure 5. Types of lattice point defects.

Linear defects generally take the form of dislocations and disclinations. Both of them were collectively referred to as "distortions" by Volterra [**25**], who first established their role in the equilibrium state of elastic media by contributing generators to the fundamental group. The dislocations amount to translations of linear frames when they are parallel-translated around a (Burgers) loop that encircles a dislocation line and can be of the edge or screw (i.e., helical) type. Hence, they have been associated with torsion [3] in the connection that defines that parallel translation, at least when one goes to a continuous distribution of dislocations [4]. Disclinations introduce a rotation of the frame around a (Frank) loop, and therefore relate to the curvature of the connection when they are continuously distributed.

Surface defects usually take the form of grain boundaries, or "walls." That is, the lattice is defined only up to that boundary surface.

One can treat the lattice defects as all defining "deletions" from the abstract simplicial complex associated with a lattice $S$.

Except for the interstitial lattice ions, which will not affect the homology, a point defect will imply deleting the vertex, in the case of vacancies and substitutions, as well as the cells that surround it when $m > 1$. In the case of a linear lattice ($m = 1$), the deletion disconnects the lattice and introduces a generator to $H_0(\Lambda_1)$. In the case of planar lattices ($m = 2$), a point defect mostly introduces a generator of $H_1(\Lambda_2)$, since it will be surrounded by a 1-cycle that no longer bounds the deleted 2-cell. In the case of a space lattice ($m = 3$), a point defect will introduce a generator of $H_2(\Lambda_3)$, since it will be surrounded by a 2-cycle that no longer bounds the missing 3-cell.

A line defect, which can only exist when $m > 1$, will disconnect a planar lattice and therefore introduce a generator of $H_0(\Lambda_2)$. When $m = 3$, it introduces a generator of both the fundamental group $\pi_1(\Lambda_3)$ and $H_1(\Lambda_3)$.

A surface defect, which can only exist when $m > 2$, disconnects a spatial lattice and thus introduces a generator of $H_0(\Lambda_3)$.

Lattice defects are not only unavoidable in real crystals, but also sometimes essential. For instance, the manufacture of semiconductors for electronics applications generally involves the intentional introduction of lattice impurities in a controlled manner, because without the impurities one would not produce the desired effect. Dislocations were introduced into the theory of the yield strength at which elastic materials give way to plastic deformation in order to account for a major discrepancy in the earlier theory. Lattice defects also play an important role in the propagation of waves of various types throughout a solid medium, such as phonons, photons, and solitons, and can lead to scattering and diffraction.

---

[3] This geometric usage of the word "torsion" is completely unrelated to its usage in the structure of Abelian groups and modules, as well as its usage in continuum mechanics.
[4] The literature on the subject of how differential geometry applies to dislocations and disclinations is vast, by now. One might confer [**26-29**] and the references cited in those books and papers.

## 2.4 Reciprocal lattices

When one embarks upon the study of wavelike excitations of crystal lattices, one finds that it is just as important to consider not only a lattice in $\mathbf{R}^n$, but also its *reciprocal lattice* in the dual space $\mathbf{R}^{n*}$.

Recall that the *dual vector space* $V^*$ to any vector space $V$ is composed of all linear functionals on $V$. If $\alpha \in V^*$ is one such functional and $\mathbf{v} \in V$ is any vector then one can denote the application of $\alpha$ to $\mathbf{v}$ by either the scalar $\alpha(\mathbf{v})$ or $<\alpha, \mathbf{v}>$. If $\{\mathbf{e}_i, i = 1, \ldots, n\}$ is a basis for $V$ then one can define a unique *reciprocal basis* $\{\theta^i, i = 1, \ldots, n\}$ by specifying that:

$$\theta^i(\mathbf{e}_j) = \delta^i_j = \begin{cases} 1 & i = j, \\ 0 & i \neq j. \end{cases} \tag{2.26}$$

Whereas a vector $\mathbf{v}$ can be expressed in the form $\mathbf{v} = v^i \mathbf{e}_i$ with respect to the chosen basis, a linear function $\alpha$ can be expressed in the form $\alpha = \alpha_i \theta^i$; in either case, we have suppressed the summation over $i$ as implicit in such cases. The evaluation of $\alpha$ on $\mathbf{v}$ then produces the number:

$$\alpha(\mathbf{v}) = a_i v^i. \tag{2.27}$$

Geometrically, one can think of the linear functional $\alpha$ as being associated with the linear hyperplane in $V$ that is defined by all vectors $\mathbf{v}$ that are annihilated by $\alpha$: i.e., all $\mathbf{v}$ such that $\alpha(\mathbf{v}) = 0$.

In the case of $V = \mathbf{R}^n$, whereas the elements of $V$ are usually represented as column vectors of real numbers, the elements of $V^*$ are represented as row matrices of real numbers.

One of the most physically important inhabitants of $\mathbf{R}^{3*}$ is the *wave covector* $k = (k_1, k_2, k_3)$ that is associated with any wave motion in $\mathbf{R}^3$. Its components $k_i = 2\pi/\lambda_i$ represent the wave numbers in each direction defined by the canonical basis $\{(1, 0, 0), (0, 1, 0), (0, 0, 1)\}$, expressed in units of radians per unit distance. Hence, when one evaluates $k$ on a position vector $\mathbf{x} = (x^1, x^2, x^3)$ the resulting number is:

$$k(\mathbf{x}) = k_i x^i = 2\pi \left( \frac{x^1}{\lambda_1} + \frac{x^2}{\lambda_2} + \frac{x^3}{\lambda_3} \right), \tag{2.28}$$

which then represents the differential increment of phase change in the direction from the origin to the point $\mathbf{x}$, as long as the $k_i$'s are constant.

The (tangent) plane annihilated by $k$ at each point is the tangent space to the constant-phase surface ($\phi$ = const.) through that point if we let:

$$k = d\theta \qquad (k_i = \partial\phi/\partial x^i). \tag{2.29}$$

The vector spaces $V$ and $V^*$ are always linearly isomorphic, at least in the finite-dimensional case. However, the choice of isomorphism is not always defined uniquely. Although this fact is crucial in geometrical matters, in which one considers tensor fields whose components must transform covariantly or contravariantly under changes of basis, it is not as essential in the name of homology. In particular, one can use the canonical basis for $\mathbf{R}^n$ and its reciprocal basis for $\mathbf{R}^{n*}$ with impunity, as well as the linear isomorphism of $\mathbf{R}^n$ with $\mathbf{R}^{n*}$ that this defines. This effectively amounts to transposing the column vector $[v^1, \ldots, v^n]^T$ to the row vector $[v^1, \ldots, v^n]$, so the components of the dual vector $\mathbf{v}^*$ to $\mathbf{v}$ with respect to the reciprocal basis $\theta^i$ will be the same as the components of $\mathbf{v}$ with respect to $\mathbf{e}_i$. This association does not, however, behave well under basis changes.

Whenever $V$ is associated with a scalar product $<.,.>$, one can also define an isomorphism of $V$ with $V^*$ by associating each $\mathbf{v} \in V$ with the linear functional $<\mathbf{v}, .>$, which takes any vector $\mathbf{w}$ to the number $<\mathbf{v}, \mathbf{w}>$. In the case of the Euclidian scalar product on $\mathbf{R}^n$, the canonical basis is orthonormal, so $<\mathbf{e}_i, \mathbf{e}_j> = \delta_{ij}$ and the components $v^i$ go to $v_i = \delta_{ij} v^j = v^i$. Thus, in this case, the isomorphism of $\mathbf{R}^n$ with $\mathbf{R}^{n*}$ that is defined by the scalar product is the same as the one obtained from associating the canonical basis with its reciprocal basis. The only difference is in the fact that any change of basis that takes the canonical basis to another orthonormal basis – i.e., any Euclidian rotation – will preserve the isomorphism of $\mathbf{R}^n$ with $\mathbf{R}^{n*}$.

For crystal lattices, the most natural basis to choose in $\mathbf{R}^m$ is the basis $\{\mathbf{a}_1, \ldots, \mathbf{a}_m\}$ that one uses to generate the Bravais lattice. For the case of $m = 3$, one can take advantage of the vector cross product on $\mathbf{R}^3$ to define the reciprocal basis $\{a^1, a^2, a^3\}$ by way of [5]:

$$a^i = \frac{1}{2V}\varepsilon^{ijk}(\mathbf{a}_j \times \mathbf{a}_k) \qquad (V = \mathbf{a}_1 \cdot (\mathbf{a}_2 \times \mathbf{a}_3)). \tag{2.30}$$

Thus, the linear functional $a^i$ annihilates the plane spanned by $\mathbf{a}_j \times \mathbf{a}_k$, while $a^i(\mathbf{a}_i) = 1$, in any case, since we have normalized the reciprocal basis by the volume $V$ of the parallelepiped spanned by $\{\mathbf{a}_1, \mathbf{a}_2, \mathbf{a}_3\}$.

When one has settled on a linear isomorphism of $\mathbf{R}^n$ with $\mathbf{R}^{n*}$, any lattice $L: S \to \mathbf{R}^n$ is automatically associated with a lattice in $\mathbf{R}^{n*}$, which one then calls the *reciprocal lattice*. That is, each lattice site corresponds to a linear functional. In particular, a Bravais lattice in $\mathbf{R}^m$ that is generated by $\{\mathbf{a}_1, \ldots, \mathbf{a}_m\}$ corresponds to a reciprocal lattice in $\mathbf{R}^{m*}$ that is generated by $\{\alpha^1, \ldots, \alpha^m\}$, so the typical reciprocal lattice site takes the form:

$$\alpha = n_k \alpha^k, \qquad (k = 1, \ldots, m), \tag{2.31}$$

---

[5] Actually, this formula is a shorthand way of describing the corresponding transformation of component matrices for $\mathbf{a}_i$ into $a^i$ with respect to the canonical bases. In practice, it often also contains a factor of $2\pi$ that we have omitted.

in which $n_k$ are the integers that generate the original lattice in $\mathbf{R}^m$.

The reciprocal unit cell that corresponds to a Wigner-Seitz cell in $\mathbf{R}^m$ is referred to as the *first Brillouin zone*. It plays an important role in the study of wave excitations in crystal lattices.

### 2.5  Cohomology and lattices

One can define a corresponding notion of linear functionals on **Z**-modules, except that the scalars that one produces are integers, not real numbers. This allows one to also define the dual **Z**-module $C^k(\Lambda_m)$ to the **Z**-module $C_k(\Lambda_m)$ and its elements are called *k-cochains* on $\Lambda_m$. The reciprocal basis to the set $\{\sigma_k(i), i = 1, \ldots, N_k\}$ of *k*-simplexes in $\Lambda_m$ is simply the set $\{\sigma^k(i), i = 1, \ldots, N_k\}$ of **Z**-linear functionals $\sigma^k(i)$ that take each simplex $\sigma_k(j)$ to the integer:

$$<\sigma^k(i), \sigma_k(j)> = \delta_{ij}. \tag{2.32}$$

A general *k*-cochain $c^k$ is then a finite formal sum:

$$c^k = \sum_{i=1}^{r_0} a_i \sigma^k(i) \tag{2.33}$$

whose coefficients $a_i$ are integers, including, possibly, zero.

Hence, if $c_k$ is a *k*-chain of the form $\sum_{i=1}^{r_k} b^i \sigma_k(i)$, where some of the $b^i$ might have to be 0 in order to sum over the same indices, then the evaluation of $c^k$ on $c_k$ is the integer:

$$<c^k, c_k> = a_i b^i. \tag{2.34}$$

One can use this bilinear pairing of $C^k(\Lambda_m)$ and $C_k(\Lambda_m)$ to define a transpose of the boundary operator in the form of $\delta^*: C^k(\Lambda_m) \to C^{k+1}(\Lambda_m)$, by way of:

$$<\delta^* c^k, c_{k+1}> = <c^k, \partial_{k+1} c_{k+1}>. \tag{2.35}$$

That is, when $\delta^* c^k$ is evaluated on any $c_{k+1}$ the resulting integer is what one obtains by evaluating $c^k$ on the boundary of $c_{k+1}$. Although this sounds abstruse, it is actually an abstraction of what one gets from Stokes's theorem, if the linear functionals on *k*-chains are defined by the integration of *k*-forms over them.

When one uses the reciprocal bases $\sigma^k(i)$ and $\sigma^{k+1}(j)$ for $C^k(\Lambda_m)$ and $C^{k+1}(\Lambda_m)$, one finds that the coboundary operator can be expressed by a matrix $[\delta^*]^{ij}$ that is simply the transpose of the incidence matrix $[\partial_{k+1}]_{ji}$ that is associated with the boundary operator, since the definition (2.35) makes $\delta^*$ the transpose – or *adjoint* – of the operator $\partial_{k+1}$.

One has analogous concepts to those of *k*-cycle and *k*-boundary in the form of *k-cocyles* and *k-coboundaries*. From (2.35), one can characterize a *k*-coboundary by the property that whenever it is evaluated on a *k*-cycle the result is zero, while a *k*-cycle has the property that whenever it is evaluated on a *k*-boundary it gives zero. The **Z**-module of $k$-cocycles is denoted by $Z^k(\Lambda_m)$, the $k$-coboundaries by $B^k(\Lambda_m)$ and the resulting quotient module $H^k(\Lambda_m) = Z^k(\Lambda_m) / B^k(\Lambda_m)$ is called the $k^{th}$ *cohomology module* of $\Lambda_m$. Hence, the elements of $H^k(\Lambda_m)$ are equivalence classes of $k$-cocycles under the equivalence relation of *cohomology*. Two $k$-cocycles $c^k$ and $c'^k$ are cohomologous if their difference is the coboundary $\delta^{k-1} c^{k-1}$ of some $k-1$-cochain $c^{k-1}$: i.e., $c^k \sim c'^k$ iff $c^k - c'^k = \delta^{k-1} c^{k-1}$.

Just as Kirchhoff's law of currents found an interpretation in terms of the homology of a network, similarly, his law of voltages has an expression in terms of cohomology. If each branch $\sigma_1(i)$ of the network is associated with a potential difference $\Delta V_i$, and $\sigma^1(i)$ is its reciprocal co-branch then the total potential difference $\Delta V = \sum_{i=1}^{r_1} \Delta V_i \sigma^1(i)$ is a 1-cochain on the network. When one evaluates it on any 1-cycle (i.e., loop) the result is the sum of the potential differences around the loop. Hence, Kirchhoff's law of voltages can be summarized by the statement that $\Delta V$ is a coboundary of some 0-cochain $V$ (which is not, of course, unique). In fact, the 0-cochain $V = \sum_{j=1}^{r_0} V_j \sigma^0(j)$ can be regarded as the electric potential at each node, so (2.35) takes the form:

$$<\Delta V, \sigma^1(i)> = <\delta^0 V, \sigma^1(i)> = <V, \partial_1 \sigma^1(i)> = [\partial_1]_{ij} V_j \qquad (2.36)$$

for any branch $\sigma^1(i)$. Hence, saying that $\Delta V$ is a coboundary is the same thing as saying that it really does represent the difference between two potentials defined at each node.

One aspect of this latter example is that we are using cohomology with coefficients in a field, namely, the real numbers. This has a big advantage in practical terms because even though the *free* **Z**-module $C_k(\Lambda_m)$ is isomorphic to the *free* **Z**-module $C^k(\Lambda_m)$ it is not generally true that the **Z**-module $H_k(\Lambda_m)$ is isomorphic to the **Z**-module $H^k(\Lambda_m)$ since they do not have to be free; however, their free summands are isomorphic. Matters simplify considerably when one goes to homology and cohomology with real coefficients, for which the **Z**-modules become **R**-vector spaces, because one is only left with the free summands, so $H^k(\Lambda_m; \mathbf{R}) = H_k(\Lambda_m; \mathbf{R})^*$.

### 2.6 Orientation and the Poincaré isomorphism

First, let us recall some manifold terminology: A *topological manifold* of dimension $n$ is a topological space such that every point has at least one neighborhood that is topologically equivalent to $\mathbf{R}^n$, or – equivalently – an open $n$-ball. The interiors supports of all geometrical simplexes fall into this category. A *topological manifold with boundary*, by comparison, is one such that every point has a neighborhood that is homeomorphic to either an open $n$-ball or its intersection with the closed half-space $\mathbf{R}^{n+}$ in $\mathbf{R}^n$ defined by – say – $x^1 \geq 0$. Examples of topological manifolds of immediate interest to us are all open subsets of $\mathbf{R}^n$, such as the interior of any simplex, and any topological space that can be triangulated into finite simplicial complex, which means it is also expressible as an $n$-chain. Other examples include spheres, torii, and projective spaces; note that none of these last three examples have boundaries as manifolds. Examples of topological manifolds with boundary are: any intersection of an open subset of $\mathbf{R}^n$ with the positive half-space $\mathbf{R}^{n+}$, such as truncations of discs and open balls, and, of course, any geometrical simplex.

However, one must be careful about distinguishing the boundary of a topological manifold and the boundary of a *triangulation* of that manifold. In the case of any simplex they are the same thing, but that is only because any simplex is *orientable*, not only in the sense that we introduced above, but also in the following sense: Suppose the n-dimensional topological manifold $M$ is triangulated by the *n*-chain:

$$[M] = \sum_{i=1}^{r_n} \sigma_n(i). \tag{2.37}$$

$M$ is *orientable* iff the $n-1$-chain $\partial_n[M]$ is a triangulation of the $n-1$-dimensional topological manifold $\partial M$; i.e.:

$$\partial_n[M] = [\partial M]. \tag{2.38}$$

Hence, the issue becomes whether one can assign appropriate orientations to the basic simplexes $\sigma_n(i)$ that make this possible or if every possible set of orientations will fail to produce this result.

In particular, a topological manifold without boundary is orientable iff one can find a triangulation of the form (2.37) such that $\partial[M] = 0$. In such a case, one then refers to $[M]$ as the *fundamental n-cycle* of $M$.

In Figure 6 we illustrate triangulations of a cylinder, which is an orientable 2-manifold with boundary, and a Möbius band, which is non-orientable, to show how these conditions apply.

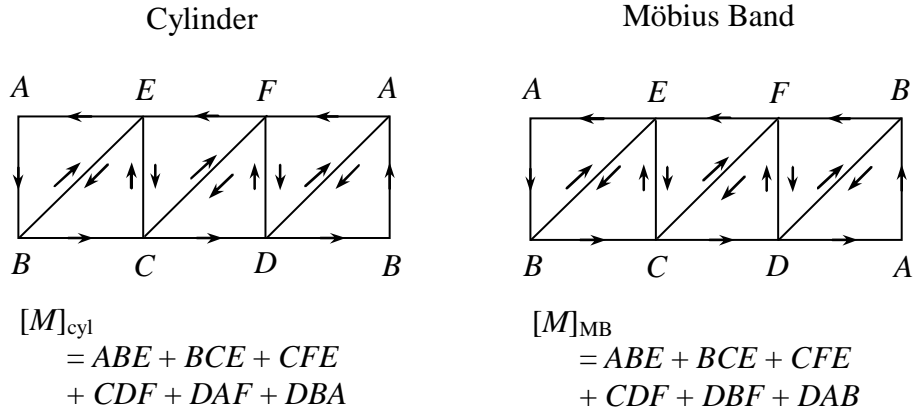

Figure 6. Triangulations of a cylinder and a Möbius band.

By direct computation, one finds that:

$$\partial[M]_{\text{cyl}} = (BC + CD + DB) - (AE + EF + FA), \tag{2.39}$$
$$\partial[M]_{\text{MB}} = (BC + CD + DA) - (AE + EF + FB) + 2AB, \tag{2.40}$$

One confirms that in the case of the cylinder the boundary of [M] represents the top and bottom triangles that remain when the two edges AB are identified, but in the case of the Möbius band the boundary of the manifold is triangulated by the 1-cycle (BC + CD + DA) + (AE + EF + FB). However, if one passes to homology "modulo 2" by replacing the coefficients of the simplexes by 0 if the coefficient is even and 1 if it is odd – keeping in mind that – $n$ is also $n$ modulo 2 – then one sees that $\partial[M]_{MB}$ agrees with the boundary of the Möbius band modulo 2, which one can then write:

$$\partial_2[M]_{MB} \cong [\partial M]_{MB} \quad (\text{mod } 2), \tag{2.41}$$

which also means that the difference between the 1-cycles on each side of the congruence is a 1-chain with even coefficients.

This is actually a special case of a more general statement that whether an $n$-dimensional manifold with boundary $M$ is or is not orientable, if it is expressible as the $n$-chain [M] then $\partial_n[M]$ triangulates $\partial M$ modulo 2: $\partial_n[M] \cong [\partial M]$ (mod 2). This is related to the fact that a non-orientable topological manifold has a two-fold covering manifold that is orientable.

As an example of a triangulation of an orientable manifold without boundary we give the 2-sphere, which we express as its homeomorphic image, the surface of a tetrahedron:

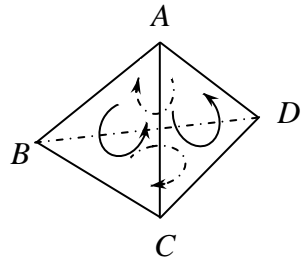

$$[S^2] = ABC + ACD + ADB + BDC$$

Figure 7. Triangulation of a 2-sphere or tetrahedron.

By direct computation, one sees that $\partial_2[S^2]$ is zero, which is consistent with one's expectation that a 2-sphere is orientable. By contrast, the projective plane $\mathbf{R}P^2$, which is obtained from the 2-sphere by identifying antipodal points, is not orientable, although verifying that $\partial_2[\mathbf{R}P^2]$ consists of a sum of 1-simplexes with even coefficients first requires a triangulation, which can be involved, so we refer the reader to the discussion in Alexandroff [17]. Of course, the two-fold covering of $\mathbf{R}P^2$ with an orientable manifold is by way of the map $S^2 \to \mathbf{R}P^2$ that identifies antipodal points.

The practical significance of orientability for us in what follows is the fact that a compact, orientable $n$-dimensional topological manifold $M$ – i.e., one that can be given a finite triangulation – will always have a generator to $H_n(M)$ in the form of the fundamental $n$-cycle [M]. (Recall that if $M$ is $n$-dimensional then the only $n$-boundary is 0.)

Furthermore, a compact orientable $n$-dimensional topological manifold $M$ also has a canonical set of isomorphisms $H_k(M) \to H^{n-k}(M)$ for each dimension $k$ that are defined by means of the fundamental $n$-cycle, or really, its dual fundamental $n$-cocycle in $H^n(M)$;

these isomorphisms are called the *Poincaré isomorphisms*. Consequently, the free and torsion parts of both modules involved have just as many generators. In particular, each path component of *M* is associated with a generator of $H^n(M)$ in the form of the fundamental *n*-cycle that triangulates it. In the event that *M* is not path-connected the fundamental *n*-cycle then decomposes into essentially independent sub-chains as a sum over all *n*-simplexes.

Closely related to the Poincaré isomorphisms, but not precisely identical to that concept, is the notion of Poincaré duality, which takes the form of isomorphisms $H_k(M) \to H_{n-k}(M)$ or $H^k(M) \to H^{n-k}(M)$ for each *k*. One needs to have isomorphisms $H_k(M) \to H^{n-k}(M)$, in addition to the aforementioned Poincaré isomorphisms, and for this, one generally needs a so-called "intersection form" on the homology modules, which works somewhat like a scalar product in the way that it defines the desired isomorphism.

We shall not go into the details of the precise definition of the Poincaré isomorphisms and duality at this point, since we shall only use them casually in the sequel. The curious are then referred to the cited literature on the topology of manifolds (e.g., [**23**, **24**]).

A consequence of Poincaré duality is that a compact, orientable, *n*-dimensional topological manifold *M* (without boundary) has its $k^{\text{th}}$ Betti number $b_k$ equal to its $(n-k)^{\text{th}}$ Betti number $b_{n-k}$. Hence, the Euler-Poincaré characteristic takes the form:

$$\chi[M] = \begin{cases} 0 & n \text{ odd} \\ \text{even integer} & n \text{ even.} \end{cases} \qquad (2.42)$$

In particular, for *n*-spheres, $\chi[S^n]$ vanishes when *n* is odd and equals 2 when *n* is even. The 2 comes about because $S^n$ is path-connected, so $b_0 = 1$ and orientable, so $b_n = 1$, by Poincaré duality, the intermediate Betti numbers vanish for spheres.

More generally, for a compact, orientable surfaces S, $\chi[M] = 2(1 - p)$, where the number *p*, which represents the number of "2-holes" is called the *genus* of the surface. For a sphere *p* is then 0, while for a torus, *p* is 1.

## 3  Order parameters

The concept of an *order field* in a material medium plays the same fundamental role in condensed-matter physics that electromagnetic fields do in the theory of electromagnetism or gravitational fields in the theory of gravitation. One should not think of the use of the word "order" as necessarily being related with the entropy density in the matter, although that concept might define a useful *example* of an order parameter. Other examples include the electric polarization vector field **P**, the magnetization vector field **M**, and the director line fields of nematic liquid crystals that define their axis of alignment without also defining the choice of two orientations that line can have.

### 3.1  Order parameter spaces

If the material medium is described by a differentiable manifold *M* of dimension *n*, which we shall call the *configuration manifold* of the medium, then the order field is described by a differentiable map $\phi: M \to R$, $x \mapsto \phi(x)$, where $R$ is an $r$-dimensional manifold called the *order parameter space*, so its elements are the *order parameters*; one also encounters the term *space of internal states* for $R$.

Although nowadays the legacy of gauge field theories is that one tends to always start with physical fields as sections of appropriate fiber bundles and regard the present construction as only being of interest locally or when the bundle is trivial, one finds that for many of the practical examples in solid-state and condensed matter physics, which tend to involve manifolds that are open subsets of $\mathbf{R}^n$, the restriction to trivial bundles is perfectly natural. One can then introduce the complicating factors that arise from the non-triviality of fiber bundles as an advanced topic, much as one starts with ideal lattices and introduces defects or starts with linear constitutive laws and introduces nonlinearities. Hence, for our present purposes, we shall discuss the simpler definition of an order field, as we have already introduced considerable complexity by way of the homology of lattices, at least as far as solid-state physics is concerned.

The introduction of homotopy into the study of order fields is associated with the concept of *topological stability*. An order field is called *topologically unstable* iff it is homotopic to a constant field. Hence, topological stability can only come about when the there is more than one homotopy class $[\phi]$ of maps from $M$ to $R$. As long as $M$ has the homotopy type of a sphere of some dimension – say, $S^n$ – the homotopy classes of order fields on $M$ can be identified with the elements of $\pi_n(R)$. This restriction is reasonable when $M$ is a vector space minus a point (the "defect") or if the order field is required to be constant at infinity, but when one has more that one defect, it becomes necessary to replace the global homotopy classes of $f$ with local ones – à la Poincaré-Hopf – that are associated with a triangulation of $M$. This then becomes the basis for obstruction theory, as we shall see.

Although many common order parameters are scalars, vectors, and tensors, which suggests that one can often use vector spaces as one's order parameter spaces, nevertheless, there are enough established examples of order parameters that belong to more general manifolds than vector spaces to justify the generalization in the eyes of physics.

For instance, when the order parameter is a quantum spin state one can use a finite set, such as $\{-\hbar/2j, -\hbar/2(j-1), \ldots, +\hbar/2(j-1), +\hbar/2j\}$, for $R$. The homotopy groups of such a set vanish except for $\pi_0(R)$, which has as $2j$ elements to it. Of course, $\pi_0(R)$ does not usually admit a natural group structure and must be treated as only a set.

When the order parameter is a scalar, vector, or tensor that is allowed to be zero, the order parameter space is simply a vector space, which is contractible. Hence, all of its homotopy groups vanish, and many of the following topological considerations are irrelevant, even for defective lattices.

However, when the order parameter is a *non-zero* vector field on the medium, one finds that since when one deletes the origin of any vector space of dimension $r+1$ the remaining topological space is homotopically equivalent to a sphere of dimension $r$ (by radial contraction), one can think of any non-zero vector field as taking its values in $S^r$, at least for the purposes of homotopy. As mentioned above, when $r > 0$ $\pi_k(S^r)$ vanishes up to $k = r$ and then $\pi_r(S^r) = \mathbf{Z}$. However, although the homotopy groups for a circle then vanish for dimensions above one, the analogous statement is not true for higher-dimensional spheres, even though it is true for homology and cohomology modules. This is one more example of how homotopy theory can be more complicated than homology. Similarly, when the order parameter is a *unit* vector field, one can also use $S^r$ for the order parameter space, as when one is specifying merely the oriented spatial direction of something.

It is also sometimes necessary to specify the *non-oriented* spatial direction of something as an order parameter, as with nematic liquid crystals. Instead of a unit vector, one must consider the entire *line* that it generates, which is homotopically equivalent to replacing the unit vector **u** with the pair {−**u**, +**u**}, which is equivalent to identifying antipodal points on the unit *r*-sphere, which defines the real projective space of *r* dimensions $\mathbf{R}P^r$. The homotopy groups of $\mathbf{R}P^r$ are then isomorphic to those of $S^r$, except for the fundamental group, which is $\mathbf{Z}_2$.

Interestingly, as far as homotopy is concerned, defining a Lorentzian metric on a manifold, such as the spacetime manifold in general relativity, is equivalent to defining a tangent line at each point of it, so the order parameter space at each point is the "projectived" tangent space – viz., the projective space of all tangent lines through the origin. Thus, the introduction of a Lorentzian metric on the spacetime manifold has much in common with the study of uniaxial nematic liquid crystals, although in the case of spacetime, it is usually harder to justify the triviality of the bundle in question, since that would make spacetime parallelizable, which is a harder condition to impose than it sounds, topologically.

Another common order parameter space with non-trivial homotopy is the *r*-dimensional torus $T^r = S^1 \times \ldots \times S^1$ (*p* factors) $= \mathbf{R}^r / \mathbf{Z}^r$. Its coordinates can usually be regarded as *r*-tuples $(\theta^1, \ldots, \theta^r)$ of rotation angles, such as phases. The homotopy groups are then isomorphic to the direct sum of *r* copies of $\pi_1(S^1) = \mathbf{Z}$ – i.e., $\mathbf{Z}^p$ – in dimension one and vanish in all higher dimensions.

The examples $S^r$, $\mathbf{R}P^r$, and $T^r$ have in common the fact all of them can be represented as *homogeneous spaces*. That is, there is some Lie group *G* with a closed subgroup *H* that makes *R* into the coset space *G*/*H*. We have already represented $T^r$ in the form $\mathbf{R}^r / \mathbf{Z}^r$, and one can always represent $S^r$ as the homogeneous space $SO(r+1) / SO(r)$, at least when $r > 1$. For instance, $S^2$ is diffeomorphic to $SO(3) / SO(2)$, while $\mathbf{R}P^r = S^r / \mathbf{Z}^2$ can be represented as $SO(r+1) / O(r)$, since the only difference between the rotations of $O(r)$ and those of $SO(r)$ is in the sign of the determinant.

The computation of the homotopy groups for a given homogeneous space usually follows from considering the long exact sequence of homotopy groups that follows from the short exact sequence of pointed spaces $H \to G \to G/H$, namely:

$$\pi_k(H) \to \pi_k(G) \to \pi_k(G/H) \to \pi_{k-1}(H) \to \pi_{k-1}(G) \to \pi_{k-1}(G/H) \to \ldots$$
$$\ldots \to \pi_1(H) \to \pi_1(G) \to \pi_1(G/H) \to \pi_0(H);$$

one calls the *homotopy exact sequence* of the fibration of *G* over *G*/*H* (see [**30**] for more details). This sequence simplifies considerably when *H* is a discrete subgroup, since in that event only $\pi_0(H)$ is non-trivial, and in fact consists of the group *H*, itself. As a result the long exact sequence breaks up into sequences of the form:

$$0 \to \pi_k(G) \to \pi_k(G/H) \to 0$$

for $k > 1$ and the final sequence:

$$0 \to \pi_1(G) \to \pi_1(G/H) \to H.$$

The advantage of the former type of short exact sequence is that it implies that $\pi_k(G)$ is isomorphic to $\pi_k(G/H)$ as a group. One can also show that the latter sequence implies that $\pi_1(G/H)$ is isomorphic to $H$. Thus, the only difference between the homotopy type of $G$ and that of $G/H$ is in dimensions zero and one. For instance, since $\mathbf{R}P^3$ is $SU(2)/\mathbf{Z}_2$, and $SU(2)$ is homotopically equivalent to $S^3$, the only thing that changes by the identification of antipodal points on the 3-sphere is that the resulting homogeneous space is no longer simply connected, but doubly connected, since $\pi_1(SU(2)/\mathbf{Z}_2)$ is isomorphic to $\pi_0(\mathbf{Z}_2)$, which is the group $\mathbf{Z}_2$.

It is no coincidence that homogeneous spaces play such a key role as order parameter spaces, since order parameter spaces usually come about as a result of spontaneous symmetry breaking. One starts with a space $X$ of "field values," such as a vector space, and a real-valued function $F: X \to \mathbf{R}$ that generally represents either the free energy in the macroscopic state of the system of atoms or molecules in the material or some tensor field that describes some physical property of the medium, such as a constitutive law. One then assumes that $F$ is symmetric under the action on $X$ of a Lie group $G$ of internal symmetries, such as rotational symmetries. Hence $F$ is constant on the orbits of that action and any orbit through a point $a \in X$ will take the form $G/G_a$, where the subgroup $G_a$ in $G$ is called the *isotropy subgroup* of the group action at $a$. In the case of the action of $SO(2)$ on $\mathbf{R}^2$ by matrix multiplication there are two possible isotropy subgroups, namely, the identity group $I$ and $SO(2)$ itself. The orbits corresponding to the former subgroup look like $SO(2)$ – i.e., the concentric circles around the origin – and the orbit whose isotropy subgroup is $SO(2)$ becomes the fixed point at the origin. Spontaneous symmetry breaking of the ground state – viz., the minimal orbit of $F$ – then takes the form of requiring that $F$ take its minimum on one of the circles, instead of the fixed point. Hence, the ground state is no longer a unique point in the internal state space, but a subset of generally higher dimension than zero.

By now, a consistent vocabulary has emerged for the topological defects in ordered media that are due to the homotopy type of $R$, and, in fact, the terms overlap the language of lattice defects somewhat. A topological defect, in any form, defines a generator of $\pi_k(R)$ in some dimension $k$. For $k = 0, 1, 2, 3$ the corresponding terms are *wall, vortex (or string), hedgehog (or monopole), and texture, resp.*

We summarize some of the most commonly studied media, their order parameters, and the first three homotopy groups in Table 1. In this table, the group $D_2$ represents the dihedral group of a rectangle, which is an Abelian group that consists of the identity transformation of $\mathbf{R}^3$ and the three inversions of each coordinate. The group $Q$ is a non-Abelian group of four elements and is isomorphic to the group of unit quaternions $\{1, i, j, k\}$ under multiplication. $E(3)$ represents the three-dimensional Euclidian group $O(3) \square \mathbf{R}^3$, where the symbol $\square$ represents the semi-direct product.

It is then important point to note that the case of non-Abelian fundamental groups is not merely a mathematical pathology in the eyes of physical examples, since it pertains to the examples in the table above of biaxial nematics and cholesterics. One finds that the non-Abelian composition of homotopy classes of loops is closely related to the issue of whether those loops are linked or not. (There is considerable discussion of the issues associated with the linking of loops in Monastrysky [**11**]).

Table 1. Common ordered media and their homotopy groups.

| Medium | Order Parameter | $R$ | $\pi_0(R)$ | $\pi_1(R)$ | $\pi_2(R)$ |
|---|---|---|---|---|---|
| Isotropic ferromagnet | magnetization vector field | $S^2$ | pt | 0 | **Z** |
| Uniaxial nematic | director line | **R**$P^2$ | pt | **Z**$_2$ | **Z** |
| Biaxial nematic | Orthonormal 3-frame (modulo discrete transformations) | $SO(3)/D_2$ | pt | $Q$ | 0 |
| Cholesteric | (same as biaxial nematic) | $SO(3)/D_2$ | pt | $Q$ | 0 |
| Smectic A | Orthonormal 3-frame (modulo discrete transformations) | $E(3)/(SO(2) \square (\mathbf{Z} \times \mathbf{R}^2))$ | pt | **Z**$_2 \square$ **Z** | **Z** |
| Smectic C | Orthonormal 3-frame (modulo discrete transformations) | $E(3)/(\mathbf{Z} \square \mathbf{Z}_2)$ | pt | **Z**$_4 \square$ **Z** | 0 |
| $^4$He II | Quantum wave function | $S^1$ | pt | **Z** | 0 |

Although $^4$He (in its II phase) has a relatively elementary structure in terms of the homotopy groups of its order parameter space, $^3$He has a very involved order parameter space, whose homotopy type depends upon various other phase parameters, one can confer the table in Appendix B of the book [**10**] by Mine'ev.

Since smectic liquid crystals have a layered structure, there has been some investigation of the possibility that one might usefully employ the methods of foliations to the modeling of their topological defects (see, e.g., [**6**]).

Since the application of homotopy methods to topological defects has been established for some time, we move on to more immediate concerns that involve the application of homology methods to defective lattices, while summarizing some of the known examples of ordered parameter spaces and their homotopy groups in Table 1.

### 3.2     Boundary-value conditions on order fields

In addition to the presence of defects in crystal lattices, generators of homology modules in various dimensions can be introduced by imposing common boundary-value conditions on the order field itself.

Suppose $L_0$ is an $m$-dimensional lattice in $\mathbf{R}^n$ that has no defects, but has a boundary $\partial L_0$. Hence, the set $L_m$ of $m$-dimensional simplexes that are generated by the vertices of $L_0$ represents a region in $\mathbf{R}^n$ that is topologically equivalent to a closed $m$-dimensional ball $B^m$, whose boundary $\partial L_m$ is topologically an $m-1$-sphere $S^{m-1}$. If one defines an order field $\phi: L_m \to R$ that must satisfy the condition that $\phi$ is constant on $\partial L_m$ then one can effectively treat $\partial L_m$ as a single point. Equivalently, when one identifies the boundary of $B^m$ to a point the resulting quotient space $B^m / \partial B^m$, whose points consist of either the interior points of $B^m$ or the boundary points, regarded as a single point, becomes topologically an $m$-sphere. This is easiest to visualize in the cases $m = 1, 2$, when one identifies the endpoints of, say, $[0, 2\pi]$ to produce a circle or one identifies the boundary of a closed disc to produce a 2-sphere, resp. The effect of this identification is then to introduce a generator for the homology module $H_m(L_m)$.

Hence, constant boundary conditions on an order field are equivalent to identifying the boundary of a lattice to a single point.

Another common boundary condition on an order field is periodic boundary conditions. For instance, if a Bravais lattice is generated by linear $m$-frame $\{\mathbf{a}_1, \ldots, \mathbf{a}_m\}$, whose lengths are $a_i$, $i = 1, \ldots, m$, resp., then one might specify that $\phi$ be periodic on the parallelepiped that they generate with spatial periods (i.e., wavelengths) in each direction that are equal to $a_i$:

$$\phi(x_1, \ldots, x_i + Na_i, \ldots, x_m) = \phi(x_1, \ldots, x_i, \ldots, x_m) \qquad (3.1)$$

for all $i = 1, \ldots, m$, $N \in \mathbf{Z}$.

The effect of periodic boundary conditions is then to identify the endpoints of each interval $(x_1, \ldots, [0, a_i], \ldots, x_m)$ of $\mathbf{R}^m$ into a circle, which then makes the resulting quotient space into an $m$-dimensional torus $T^m = \mathbf{R}^m / \mathbf{Z}^m = S^1 \times \ldots \times S^1$. Topologically, this has the effect of introducing $m$ independent generators for $H_1(L_m)$.

### 4     The continuum limit

Admittedly, in this day and age it is more physically fundamental to regard any real-world material as ultimately composed of atoms, whether isolated or bound together into molecules or lattices of ions. Nevertheless, the samples of materials that are used in laboratory experiments generally involve enormous numbers ($> 10^{23}$) of atoms that only resolve to a discrete structure at some scale of distance that might lie beyond direct measurement, except when one goes to quantum techniques such as X-ray, neutron, or electron diffraction. Thus, it is often still useful to model the macrostates of physical systems by means of distributions and fields defined on continuous regions of space.

This process of passing from the discrete system of microstates, such as a volume of gas molecules or a crystal lattice, to a corresponding system of macrostates that refer to a continuous region of space is generally referred to as passing to the *continuum limit*. In order to make it a limit in the analytical sense of the word, one can think of the process as involving increasing the total number $N$ of material points in the region without bound, while maintaining a constant average density of the quantity in question over the total volume $V$. This is distinct from the *thermodynamic limit*, in which one also allows $V$ to become infinite, as well.

The case at hand begins with a lattice $L: S \to \mathbf{R}^n$, as above, whose image then becomes a finite subset of $\mathbf{R}^n$ that we denote by $L_0$. We shall be interested in the process of extending an order field $\phi: L_0 \to \Pi$ to a continuous field defined on "the space spanned by $L_0$." Of course, the latter phrase in quotes needs to be made more rigorous, so we treat it as meaning that if $S$ is associated with an abstract simplicial complex whose vertex scheme is defined by the points of $S$ themselves then $L_0$, which we now denote by $\Lambda_0$, is the 0-skeleton of its geometrical realization by points in $\mathbf{R}^n$, $L_k$ is its set of geometrical realizations of the $k$-simplexes, and, in general, $\Lambda_k$ is the $k$-skeleton of that geometrical realization. Ultimately, if $S \subset \mathbf{Z}^m$ then the $m$-skeleton $\Lambda_m$ of the complex is a polyhedral region in $\mathbf{R}^n$ that will serve to give a precise meaning to the phrase "the space spanned by the lattice."

In one sense, the problem of extending a field defined on a lattice to a field defined on a continuous region that contains it is a problem of interpolation. For some intriguing insights into how one might approach the problem from that angle in practice, one might confer the paper of Krumhansl [**31**], in which he applies a technique that is borrowed from communications theory and is based on a theorem of Claude Shannon.

Increasing $N$ to infinity can take the form of making successively finer subdivisions of the lattice by the interpolation of more lattice sites. This implies corresponding subdivisions of the associated simplicial complex, which must be of the type that preserves the homology modules, up to isomorphism, such as barycentric subdivisions of triangular complexes.

In the context of topology, the problem of extending the order field $\phi : \Lambda_0 \to R$ from a lattice to a continuous order field $\phi_m : \Lambda_m \to R$ falls within the purview of the most elementary form of obstruction theory, which we now discuss.

## 5  Obstructions to continuous extensions

This problem that we just posed is usually first discussed in algebraic topology in the more general form of finding topological obstructions to the continuous extension of a map $f: A \to Y$, when $A \subset X$ is a subset of one topological space $X$ and $Y$ is another topological space, to a map $f: X \to Y$ (see [**15**, **30**, **32, 33**] on this).

The usual approach to the continuous extension problem is to first assume that $X$ is represented as the geometrical realization of a simplicial complex $K$ (actually, one usually encounters *CW-complexes*, whose basic building blocks are based on balls, but the effect on homology is the same) and $A$ is a subcomplex of that complex. One begins with $f$ being defined on the 0-skeleton $K_0$ of the complex, extending to the other points outside of $A$ using an arbitrary constant element of $Y$, if necessary. One then proceeds stepwise by examining the extension to the 1-skeleton, 2-skeleton, etc.

In general, if $b_k$ is the boundary $\partial\sigma_{k+1}$ of a $k+1$-simplex $\sigma_{k+1}$ then since $\sigma_{k+1}$ is contractible to a point an extension of $f$ from $\partial\sigma_{k+1}$ to $\sigma_{k+1}$ is possible iff the homotopy class $[f(\partial\sigma_{k+1})]$ of $f: \partial\sigma_{k+1} \to Y$ is trivial. Since $\partial\sigma_{k+1}$ is homeomorphic to a a $k$-sphere, $[f(\partial\sigma_{k+1})] \in \pi_k(Y)$. This association of each $\sigma_{k+1}$ with $[f(\partial\sigma_{k+1})] \in \pi_k(Y)$ is then extended by linearity to a $k+1$-chain $c_{f,k}$ on $X$ with values in the group $\pi_k(Y)$. By this, we mean that if:

$$c_{k+1} = \sum_{i=1}^{N} m_i \sigma_{k+1}(i) \tag{5.1}$$

then

$$\langle c_{f,k}; c_{k+1}\rangle = \sum_{i=1}^{N} m_i [f(\partial\sigma_{k+1}(i))]. \tag{5.2}$$

This makes $c_{f,k}$ an element of $C^{k+1}(X; \pi_k(Y))$, so $\langle c_{f,k}; c_{k+1}\rangle$ is an element of $\pi_k(Y)$.

In fact, $c_{f,k}$ is a $k+1$-cocycle $c_{f,k} \in H^{k+1}(X; \pi_k(Y))$, although in the case of $k = 0$, one must recall that $\pi_0(Y)$ generally has no natural group structure and for $k = 1$, one might have to use the Abelianization of $\pi_1(Y)$. The fact that $c_{f,k}$ is a cocycle follows from the fact that if $b_{k+1}$ is $k+1$-boundary then it can be expressed in the form:

$$b_{k+1} = \sum_{i=1}^{N} m_i \partial\sigma_{k+2}(i), \tag{5.3}$$

which makes:

$$\langle c_{f,k}; b_{k+1}\rangle = \sum_{i=1}^{N} m_i [f(\partial^2 \sigma_{k+2}(i))] = 0. \tag{5.4}$$

Thus, when $c_{f,k}$ is evaluated on any boundary, it vanishes, and this, as we pointed out above, is one way of characterizing cocycles.

The necessary and sufficient condition for the continuous extension of $f$ from $K_k$ to $K_{k+1}$ is then the vanishing of $c_{f,k}$.

The first non-vanishing $c_{f,k}$ as one goes from $k = 0$ upwards is referred to as the (primary) *obstruction cocycle* of $f$. One sees that $H^{k+1}(X; \pi_k(Y))$ can be non-zero iff both $H^{k+1}(X)$ and $\pi_k(Y)$ are non-vanishing at the same time. Thus, $c_{f,k}$ will automatically vanish whenever either $H^{k+1}(X)$ or $\pi_k(Y)$ are trivial. For instance, if $V$ is a vector space then any $f: A \subset X \to V$ can be continuously extended to $X$ since $\pi_k(V) = 0$ for all $k$. Similarly, any constant map on $A$ can be continuously extended to a constant map on $X$.

## 6   Physical examples

Let us now apply this process to the case at hand of extending an order field $\phi: \Lambda_0 \to R$ from a lattice $\Lambda_0$, which we regard as the 0-skeleton of an $m$-dimensional polyhedral complex $\Lambda_m$, to a continuous map $\phi: \Lambda_m \to R$. Thus, we shall be concerned with elements $c_{\phi,k}$ of $H^{k+1}(\Lambda_m; \pi_k(R))$, $k = 0, \ldots, m$, which makes it clear that when the lattice and the order parameter space are both topologically defective there could be topological obstructions to this extension.

Since the dimension $m$ of the lattice is usually 1, 2, or 3, in practice, this means that we shall need to consider only $H^1(\Lambda_m; \pi_0(R))$, $H^2(\Lambda_m; \pi_1(R))$, and $H^3(\Lambda_m; \pi_2(R))$. Therefore, textures (which belong to $\pi_3(R)$) would not be an issue until one considered a four-dimensional lattice.

In each case, we will examine how the obstruction cocycle in that dimension works for various physical examples of order fields on defective linear, planar, and space lattices when the order parameter space also has topological defects of the appropriate dimension.

### 6.1    1-cocycle on a lattice with values in $\pi_0(R)$

As mentioned above, the case of $k = 0$ requires special attention, since $\pi_0(R)$ is a usually a set, not a group. Of course, when $R = G/H$ is a homogeneous space with a discrete subgroup $H$, $\pi_0(R)$ will indeed be a group. Also, since the ultimate result of the analysis is merely that an order field on a 0-skeleton of a lattice can be continuously extended to the 1-skeleton iff it is constant on the path components of the 0-skeleton, one sees that the only benefit to going through the argument naively is simply to illustrate the significance of obstruction cocycles in an elementary setting.

The non-vanishing of $H^1(\Lambda_m; \pi_0(R))$ would imply that the lattice was multiply connected while the order parameter space was not path connected. Thus, a linear lattice would have to be a finite set of disconnected non-simply connected sub-lattices, a planar lattice would have to have point defects or be orientable and without boundary, while a space lattice would have to have dislocations. The appearance of 1-cycles in a linear lattice might follow from imposing boundary-value conditions on the order field. A typical example of an order parameter space that is not path-connected might be the quantum spin space $R = \{-\hbar/2, +\hbar/2\}$, for which we shall denote the elements of $\pi_0(R)$ by $-1$ and $+1$.

First, consider a path-connected linear lattice $\Lambda_1$ for which $H^1(\Lambda_1)$ is non-trivial; when there is more than one path component to $\Lambda_1$, one simply repeats the argument for each separate component.

Let the fundamental 1-cycle on $\Lambda_1$ be represented as:

$$[\Lambda_1] = \sum_{i=1}^{N_1} \sigma_1(i), \tag{5.5}$$

where $N_1 = N_0 + 1$ represents the number of branches that connect the vertices.

Now, let $\phi: \Lambda_0 \to R$ be an order field that is initially defined on the vertices of the lattice and takes its values in an order parameter space $R$ with $\pi_0(R)$ non-trivial.

The obstruction cocycle $c_{\phi,0}$, when evaluated on $[\Lambda_1]$, gives:

$$<c_{\phi,0}; [\Lambda_1]> = \sum_{i=1}^{N_1} [\phi(\partial \sigma_1(i))]. \tag{5.6}$$

Of course, when $\pi_0(R)$ is not a group the addition of terms is not rigorously defined. Similarly, the winding number of f is not defined, so we simply define for our purposes: Let $\partial\sigma_1(i) = \sigma_{0,1}(i) - \sigma_{0,0}(i)$. Define:

$$[\phi(\partial\sigma_1(i))] = \begin{cases} 0 & \text{if } \phi \text{ maps } \sigma_{0,0}(i) \text{ and } \sigma_{0,0}(i) \text{ into the same path component of } R \\ 1 & \text{otherwise.} \end{cases}$$

Hence, we interpret the vanishing of the expression (5.6) to mean that all individual terms are zero. Thus, since $[\Lambda_1]$ is a cycle, each vertex that ends one 1-simplex begins another one, and if $<c_{\phi,0}\,;[\Lambda_1]>$ vanishes then $\phi$ must be constant over all vertices. The obstruction to extending a quantum spin field from the vertices to the loop is then the constancy of its values.

Similar arguments apply to the case of planar lattices with point defects and space lattices with dislocations. The loop then becomes a non-bounding 1-cycle.

### 6.2    2-cocycle on a lattice with values in $\pi_1(R)$

Now, let us look at the obstruction cocycle in $H^2(\Lambda_m; \pi_1(R))$. The non-vanishing of that cohomology module would imply that the $H^2(\Lambda_m)$ and $\pi_1(R)$ were both non-vanishing. In the former case, one could be either dealing with an orientable planar lattice or a space lattice with point defects, while the order parameter space would have to have a vortex defect. Possible candidates for $R$ might be $S^1$ or $\mathbf{R}P^n$, for which $\pi_1(R)$ would be isomorphic to $\mathbf{Z}$ and $\mathbf{Z}_2$, respectively.

If we represent a typical 2-cycle in the form:

$$z_2 = \sum_{i=1}^{N_2} a_i \sigma_2(i) \tag{5.7}$$

then the obstruction cocycle $c_{\phi,1}$, when evaluated on $z_2$, gives:

$$<c_{\phi,1}\,;z_2> = \sum_{i=1}^{N_2} a_i [\phi(\partial\sigma_2(i))]. \tag{5.8}$$

In the case of $R = S^1$, for which $\pi_1(R) = \mathbf{Z}$, this expression will be a sum of integers $[\phi(\partial\sigma_2(i))]$ that represent the winding numbers of the maps $\phi: \partial\sigma_2(i) \to R$ that are defined by the restriction of $\phi$ to the boundary of each 2-simplex. When In the case of $R = \mathbf{R}P^n$, for which $\pi_1(R) = \mathbf{Z}_2$, each $[\phi(\partial\sigma_2(i))]$ will be 0 or 1 depending upon whether the winding number is even or odd, respectively. Similarly, the sum will be 0 or 1 depending upon whether it is even or odd, as well. Thus, in either case, one sees that the vanishing of $c_{\phi,1}$ does not have to imply that the winding numbers of $\phi$ on each face $\partial\sigma_2(i)$ vanish, but only their sum over all faces.

A historically important interpretation of this situation in obstruction theory is the closely-related problem of finding non-zero *tangent* vector fields on compact differentiable manifolds. Although this actually requires passing to "local systems of homotopy groups" because the space $R$ is now a tangent space to a potentially non-parallelizable manifold, many of the same ideas we have dealing with all along assert themselves.

If a tangent vector field **v** on a manifold $M$ has a zero at a point $x_0$ then one defines the *index* of that zero Ind[$x_0$] to be the winding number [**v**($\partial\sigma_n(x_0)$)] of the map **v**: $\partial\sigma_n(x_0) \to \mathbf{R}^n - \{0\}$, where $\sigma_n(x_0)$ is a sufficiently small disc surrounding $x_0$ and, as one recalls, $\mathbf{R}^n - \{0\}$ is homotopically equivalent to an $n-1$-sphere. Thus, as long as $\sigma_n(x_0)$ is part of a larger triangulation of $M$ into the fundamental $n$-cycle, we can say that:

$$<c_{\mathbf{v},n-1} ; [M]> = \sum_{\text{all zeroes}} \text{Ind}[x_0]. \tag{5.9}$$

A deep and ground-breaking theorem of topology that was first proved for surfaces by Poincaré and later extended to compact differentiable manifolds by Hopf, is that this sum also equals $\chi[M]$. Thus, the obstruction $n$-cocycle $c_{\mathbf{v},n-1}$, which has its values in $\mathbf{Z}$, is also referred to as the *Euler class* of $M$, or really, its tangent bundle. A compact differentiable manifold admits a global non-zero vector field iff its Euler-Poincaré vanishes, so, in particular, all compact, orientable manifolds of odd dimension admit non-zero vector fields, while many even-dimensional manifolds do not.

The canonical example of a compact surface that does not admit a global non-zero tangent vector field is the 2-sphere, for which for which $H^0(T^2) = \mathbf{Z}$, since it is path connected, $H^2(T^2) = \mathbf{Z}$, since it is compact and orientable, which is also consistent with Poincaré duality, and $H^1(T^2) = 0$, since it is simply connected. The obstruction to continuously extending a non-zero tangent vector field that is defined on the 1-skeleton of a triangulation of $S^2$ to a non-zero tangent vector field on $S^2$ is then a 2-cocycle $c_{\mathbf{v},1}$ with values in the integers, which, when evaluated on the fundamental 2-cycle $[S^2]$ gives $\chi[S^2] = 1 - 0 + 1 = 2$. An example of a vector field on the 2-sphere is any vector field that is tangent to the longitude circles and points south, which then has zeroes at the poles. Despite the fact that the vector field is radial outward at the North pole and radial inward at the South pole, its index at each is + 1. One can define an example of a tangent vector field with a single zero of index 2 by intersecting the sphere with all planes through a fixed line tangent to the North pole.

An even-dimensional example of a manifold that does admit a non-zero vector field is the 2-torus $T^2 = S^1 \times S^1$, for which $H^0(T^2) = \mathbf{Z}$, since it is path connected, $H^2(T^2) = \mathbf{Z}$, since it is compact and orientable, which is also consistent with Poincaré duality, and $H^1(T^2) = \mathbf{Z} \oplus \mathbf{Z}$, since $H^1(T^2)$ is isomorphic to $\pi_1(T^2) = \pi_1(S^1) \oplus \pi_1(S^1)$, in this case, which gives $\chi[T^2] = 1 - 2 + 2 = 0$. Hence, a 2-torus always admits global non-zero tangent vector fields, and, indeed, any $n$-torus.

Note that the Poincaré-Hopf theorem does not generally apply in the case of vector fields that are not necessarily tangent to submanifolds in $\mathbf{R}^n$, since a 2-sphere can admit a constant vector field when one drops the restriction that the vectors be tangent to it.

As a first physical example of the previous remarks, let us consider a non-zero quantum wave function $\phi: \Lambda_2 \to \mathbf{C} - \{0\}$ on an orientable planar lattice $\Lambda_2$ that we assume to close into a spherical lattice. We might simply be imposing constant boundary conditions, or perhaps a "Bucky-ball" might serve to illustrate such a lattice in material reality. Thus, $H^2(\Lambda_2)$ will have one generator, in the form of the fundamental 2-cycle $[\Lambda_2]$, while $\pi_1(R)$ is $\mathbf{Z}$ since the homotopy of the punctured plane is carried by the unit circle. If the values of $\phi$ are represented in polar form as $Re^{i\theta}$ then it is sufficient consider the circular part $e^{i\theta}$ as far as homotopy is concerned.

Note that since we are not requiring that the plane of $\mathbf{C}$ be tangent to the sphere in question, the Poincaré-Hopf theorem cannot be used.

Since winding numbers can take on all positive and negative values, it is clear that the vanishing of the right-hand side of (5.8) does not have to imply that all of the terms vanish, only that the sum of the positive contributions equals the sum of the negative contributions. Obviously, if $\phi$ is constant on the 1-skeleton $\Lambda_1$ then all of the winding numbers vanish and one can extend $\phi$ continuously to $\Lambda_2$. However, non-constant fields can still have non-zero winding numbers that sum to zero.

The same considerations apply to the case of a non-zero quantum wave function on a planar lattice with periodic boundary conditions or a space lattice with point defects.

In the case of a director field on a space lattice with point defects, we shall assume that the directors are lines in the ambient space $\mathbf{R}^3$, so as to avoid the consideration of non-trivial bundles. Thus $H^2(\Lambda_3)$ has as many generators as defects and $\pi_1(R) = \pi_1(\mathbf{RP}^2) = \mathbf{Z}_2$, which we regard as composed of the set {even, odd} when given addition. In the case of loops in $\mathbf{RP}^2$, it is convenient to imagine them as consisting of rotations of lines through the origin of $\mathbf{R}^3$ around a circle in some plane through the unit sphere, such as the equatorial plane or a polar plane. One can then illustrate the two possible loops as follows:

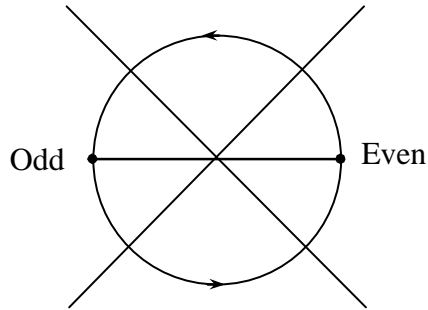

Figure 8. Schematic depiction of the winding number for loops in projective spaces.

In interpreting this diagram, we intend that rotating a horizontal line through the origin through an odd multiple of $\pi$ will close a loop in $\mathbf{RP}^2$ into one homotopy class, while rotating it through an even multiple of $\pi$ will close it into a loop of the other class.

One might note that since the homotopy groups of spheres and projective spaces differ only in dimension one, the only time the continuous extensions of non-zero vector fields and line fields (i.e., director fields) are obstructed by different cocycles is in the case where $H^2(\Lambda_m)$ is non-trivial.

### 6.3 3-cocycle on a lattice with values in $\pi_2(R)$

In order for $H^3(\Lambda_m)$ to have a non-zero generator, first, we would clearly have to have a lattice whose dimension was at least three. If we were indeed concerned with a space lattice then since there are no 3-boundaries in three dimensions $H^3(\Lambda_3) = Z^3(\Lambda_3)$ and a non-zero generator of $H^3(\Lambda_3)$ would take the form of a non-zero 3-cycle. Since the lattices that we have in mind are generated by finite sets of vertices their $k$-skeletons are all compact. As a result, the only way that $H^3(\Lambda_3)$ can vanish is if $\Lambda_3$ is not orientable, such as one finds with a "Möbius crystal." However, if $\Lambda_3$ is orientable then one always finds that each path component of it is associated with a non-vanishing 3-cocycle, as a result of Poincaré isomorphism, in the form of the fundamental 3-cocycle:

$$[\Lambda_3] = \sum_{i=1}^{N_3} \sigma_3(i). \tag{5.10}$$

Imposing constant boundary conditions on $\phi$ when $\Lambda_3$ has no defects would give a triangulation of $S^3$, while periodic boundary conditions would give $T^3$. In either case, $H^3(\Lambda_3) = \mathbf{Z}$.

For $\pi_2(R)$ to be non-trivial, one would have to be considering an order parameter space with point defects, such as $S^2$ or $\mathbf{R}P^2$, which have the same $\pi_2(R)$, namely, $\mathbf{Z}$.

The obstruction to extending an order field $\phi$ from $\Lambda_2$ to $\Lambda_3$ is a 3-cocycle $c_{\phi,2}$ with values in the integers, which, when evaluated on the fundamental 3-cycle, gives:

$$<c_{\phi,2}\,; [\Lambda_3]> = \sum_{i=1}^{N_3} [\phi(\partial \sigma_3(i))]. \tag{5.11}$$

Again, if the order field is a non-zero tangent vector field on a compact 3-manifold $M$, then since $\chi[M] = 0$ in any case, the obstruction to continuously extending a non-zero tangent vector field from $\Lambda_2$ to $\Lambda_3$ is the Euler class, and its value on the fundamental 3-cycle always vanishes.

Similarly, if the vector field is not required to be tangent then Poincaré-Hopf does not apply.

### 7 Discussion

To summarize what we established in the preceding study:
1. A crystal lattice defines an abstract simplicial complex and its geometric realization in the space of the crystal.
2. Lattice defects introduce generators of the integer homology modules in each dimension.
3. When an order field takes its values in an order parameter space with non-trivial homotopy groups, it is possible that there are non-vanishing cocycles in the cohomology of the lattice that take their values in the homotopy group of one lower dimension.

4. These cocycles represent obstructions to the continuous extension of order fields defined on lower-dimensional skeletons of the simplicial complex of the lattice to the skeleton of next higher dimension.

5. Passing to the "continuum limit" of an infinitude of infinitesimally close lattice sites will then be obstructed by these cocycles. Thus, the continuum limit of the order field will not be defined at some set of "singular" points.

One of the topics we implicitly left out above was based in the fact that we were using coefficient groups, namely, the groups $\pi_k(R)$, that were Abelian. Since non-Abelian $\pi_1(R)$ play an important role in the established theory of topological defects in order media, this must be addressed eventually. One gets into the deep issues of homology theory involving the consideration of "linking coefficients." For instance, suppose one thinks of two loops in $M$ as two 1-cycles. If one of them bounds a 2-chain then one can assign an integer to the number of times the other intersects that 2-chain – up to orientation – as a measure of the extent to which the loops are linked together.

In the foregoing discussion, we also summarily ignored the consideration of the boundary to the lattice by considering only order fields that obeyed appropriate boundary conditions that would make such a consideration unnecessary. Since there are such things as topological defects in ordered media that can manifest themselves on surfaces, such as "boojums" in superfluid helium, one then must move to the study of the *relative* homology modules of the lattice. A *k*-chain in a topological space X is called a *relative k-cycle* relative to a closed subspace *A* iff its boundary is a *k*-chain in *A*. Two relative *k*-cycles in *X* are homologous relative to *A* iff their difference is a relative *k*-boundary; that is, a *k*-chain in *X* that differs from a boundary in *X* by a chain in *A*.

One then must consider topological obstructions as relative cocycles in *M* (modulo $\partial M$) with values in the homotopy groups of *R*.

As mentioned above, if one wishes to generalize the order fields to sections of non-trivial fiber bundles whose fibers are homotopically equivalent to *R* then one must also replace the homotopy groups of *R* with a "local system" of homotopy groups for the fibers of the bundle. This gets one into the much-discussed realm of characteristic classes for fiber bundles, which has been a common focus in gauge field theories. For some remarks on how that would apply to spacetime defects, one might confer Delphenich [**33**].

One of the most fundamental topics relating to solid-state and condensed matter physics is the propagation of waves through such media, in the form of acoustic, electromagnetic, and matter waves. As we saw above, the concept of a reciprocal lattice relates to wave covectors, on the physics side of the issue, and cohomology, on the mathematical side. One then suspects that cohomology might have much to say about the propagation of waves in defective lattices. Indeed, in practice the very existence of lattice defects is often deduced from the way that waves propagate.

Other directions of extension for the methods discussed above might include lattices of dimension higher than three, such as one encounters in lattice gauge theories, which also considers the continuum limit, and the meshes used in the numerical models for the solutions of systems of differential equations, such as in finite-element analysis, or numerical relativity.